\documentclass[%
reprint,
superscriptaddress,
 amsmath,amssymb,
prb,
onecolumn,
]{revtex4-2}

\pdfoutput=1

\usepackage{graphicx}
\usepackage{hyperref}

\begin{document}


\title{Devolatilization of Subducting Slabs, Part I: \\
Thermodynamic Parameterization and Open System Effects}


\author{Meng Tian}
\email[]{meng.tian@csh.unibe.ch}
\altaffiliation[Now at: ]{Center for Space and Habitability, University of Bern, Bern, Switzerland}
\affiliation{Department of Earth Sciences, University of Oxford, South Parks Road, Oxford, OX1 3AN, UK.}

\author{Richard F. Katz}
\affiliation{Department of Earth Sciences, University of Oxford, South Parks Road, Oxford, OX1 3AN, UK.}

\author{David W. Rees Jones}
\affiliation{Department of Earth Sciences, University of Oxford, South Parks Road, Oxford, OX1 3AN, UK.}
\affiliation{Department of Earth Sciences, Bullard Laboratories, University of Cambridge, Madingley Road, Cambridge, CB3 0EZ, UK.}
\affiliation{School of Mathematics and Statistics, University of St Andrews, North Haugh, St Andrews, KY16 9SS, UK}



\begin{abstract}
The amount of H$_2$O and CO$_2$ that is carried into deep mantle by subduction beyond subarc depths is of fundamental importance to the deep volatile cycle but remains debated. Given the large uncertainties surrounding the spatio-temporal pattern of fluid flow and the equilibrium state within subducting slabs, a model of H$_2$O and CO$_2$ transport in slabs should be balanced between model simplicity and capability. We construct such a model in a two-part contribution. In this Part I of our contribution, thermodynamic parameterization is performed for the devolatilization of representative subducting materials---sediments, basalts, gabbros, peridotites. The parameterization avoids reproducing the details of specific devolatilization reactions, but instead captures the overall behaviors of coupled (de)hydration and (de)carbonation. Two general, leading-order features of devolatilization are captured: (1) the released volatiles are H$_2$O-rich near the onset of devolatilization; (2) increase of the ratio of bulk CO$_2$ over H$_2$O inhibits overall devolatilization and thus lessens decarbonation. These two features play an important role in simulation of volatile fractionation and infiltration in thermodynamically open systems. When constructing the reactive fluid flow model of slab H$_2$O and CO$_2$ transport in the companion paper Part II, this parameterization can be incorporated to efficiently account for the open-system effects of H$_2$O and CO$_2$ transport.
\end{abstract}


\maketitle

\section{Introduction}
Subduction zones are sites where two tectonic plates converge and one descends beneath the other. As subduction proceeds, the warm ambient mantle heats the downgoing plate, which sinks to increasing depth and pressure. The thermal and mechanical changes drive chemical reactions in the slabs, among which devolatilization reactions are of particular importance because they induce flux melting, trigger seismic activity, and mobilize diagnostic chemical species observed in arc lava chemistry \citep[see][for a review]{Schmidt:2014aa}. When it comes to global volatile cycling, especially H$_2$O and CO$_2$, controversy persists regarding where and how much H$_2$O \& CO$_2$ release occurs in subducting slabs \citep{Grove:2012aa, Dasgupta:2013aa, Gorman:2006aa, Kelemen:2015aa}. For example, \citet{Dasgupta:2013aa} and \citet{Gorman:2006aa} argue that subduction brings more CO$_2$ into the deep mantle than the amount emitted from arc volcanoes, based respectively on experiments and thermodynamic modelling. However, after compiling CO$_2$ flux data from field studies and extrapolating modelling results, \citet{Kelemen:2015aa} contends that almost all the subducted CO$_2$ is returned to the overriding lithosphere or surface during subduction. Studies focusing only on H$_2$O transport in subduction zones also yield disparate conclusions with regard to the efficiency of H$_2$O recycling into the deeper mantle: \citet{Hacker:2008aa} and \citet{Keken:2011aa} suggest that 35--54\% of subducted H$_2$O is recycled to mantle depths beyond $\sim$150 km, whereas \citet{Wada:2012aa} argues that only half of this amount is recycled if the subducting slab is heterogeneously hydrated as opposed to uniformly hydrated.

Large uncertainties surround the attempt to quantify the H$_2$O and CO$_2$ fluxes within and from subducting slabs. On the dynamic side, the style and pathways of volatile migration is poorly constrained. Field evidence shows that fluid flow proceeds through both porous and fractured rocks \citep[see][for a review]{Ague:2014ab} and numerical models suggest that fluid migration can take place via porosity waves \citep{Morishige:2018aa}, fractures \citep{Plumper:2017aa}, and large-scale faults \citep{Faccenda:2009aa}. Recent geochemical studies indicate that fluid explusion during subduction might be episodic rather than steady \citep{John:2012aa}. On the thermodynamic side, although it is known that metamorphic devolatilization reactions control the release of slab H$_2$O and CO$_2$, it remains unclear whether or not these reactions achieve equilibrium or overstep it \citep[e.g.][]{Hetenyi:2007aa, Dragovic:2012aa, Castro:2017aa}. Moreover, earlier experimental work \citep{Watson:1987aa} shows that the dihedral angle between mineral grains coexisting with released volatiles is highly variable and can prevent the formation of an inter-connected pore network during subduction.

On top of the uncertainties above, from a modelling perspective, a computational obstacle in quantifying the H$_2$O \& CO$_2$ budget in subducting slabs comes from thermodynamically open-system behavior. Volatiles that are released during devolatilization reactions are buoyant relative to surrounding rocks and tend to migrate \citep[e.g.,][]{Plumper:2017aa}. As conventional thermodynamic computation for chemical equilibrium assumes closed systems \citep[e.g.,][]{Connolly:2002aa, Powell:1998aa}, the open-system behavior caused by volatile migration makes it challenging to model the slab H$_2$O \& CO$_2$ budget. To circumvent this challenge, \citet{Gorman:2006aa} approximate the effects of mass transfer by adjusting the bulk H$_2$O \& CO$_2$ content of the modelled rocks according to simple rules inspired by buoyancy-driven flow. The adjustment is premised on pre-defined fluid flow pathways, along which upstream H$_2$O \& CO$_2$ is added to downstream bulk compositions and re-equilibration is repetitively calculated. Such an approach is advantageous in providing all the thermodynamic information of the modelled system, e.g., mineral mode, composition, density, etc., but suffers from oversimplification in assuming upward fluid migration. In contrast, fluid dynamical calculations \citep{Wilson:2014aa} show a substantial up-slab fluid flow during subduction. Another way of treating open-system behaviors comes from studies of magma dynamics where percolation of partially molten melts causes mass transport \citep[e.g.,][]{Katz:2008aa, Keller:2016aa}. The essence of this approach lies in the simplification of thermodynamics, such that the computational cost of thermodynamic calculation becomes affordable when it is coupled with fluid dynamic simulations. Its advantage is therefore a consistent treatment of the flow and reactions. 

However, given the large uncertainties regarding the dynamics and thermodynamics of volatile transport in subducting slabs, it is important to focus on robust, leading-order phenomena in constructing models to evaluate the H$_2$O \& CO$_2$ budget and fluxes in the slabs. The leading-order factors in our consideration are: the coupling between (de)hydration and (de)carbonation reactions, open-system behaviors caused by volatile transport, and the direction of volatile migration. We handle the coupled dehydration and decarbonation reactions with a thermodynamic parameterization that is amenable to systems open to H$_2$O \& CO$_2$ transport during subduction. In the companion contribution (Part II), we apply the parameterization to subduction zones to assess the effects of open-system behavior and fluid flow directions on slab H$_2$O \& CO$_2$ budget and fluxes. The simplified calculations introduced by the parameterization overcome the obstacle of computational cost of this coupling.

In the following, we first present in section \ref{sec:strategy} the strategy and formalism adopted to parameterize the subduction-zone devolatilization reactions. The strategy is then applied separately to each representative subducting lithology in section \ref{sec:results}. In this process, our parameterization ensures two leading-order features: one is that the CO$_2$ content of the liquid phase increases with temperature and decreases with pressure, as confirmed by previous experiments \citep[e.g.,][]{Molina:2000aa}; the other is that preferential H$_2$O loss from or CO$_2$ addition to the bulk system raises the onset temperature of devolatilization and thus inhibits it. Through simple examples in section \ref{sec:examples}, we show that inclusion of these two features enables the parameterization to simulate fractionation and infiltration processes relevant to open-system behaviors \citep[e.g.,][]{Gorman:2006aa}. An understanding of fractionation and infiltration is crucial for interpreting the results on the H$_2$O \& CO$_2$ storage and fluxes in the slabs that are presented in the companion manuscript. In section \ref{sec:discussion}, the limitations of this thermodynamic parameterization are discussed. 

\section{Strategy and formalism} \label{sec:strategy}
The goal of our thermodynamic parameterization is, for a specified pressure ($P$), temperature ($T$), and bulk H$_2$O \& CO$_2$ content of a given lithology, to predict the quantitative equilibrium partitioning of H$_2$O \& CO$_2$ between solid rock phase and liquid volatile phase. Such a thermodynamic module is therefore applicable to a system that comprises two physical phases of liquid and solid, and multiple chemical components of $\mathrm{Na_2O}$, $\mathrm{CaO}$, $\mathrm{SiO_2}$, H$_2$O, CO$_2$, etc. To make the parameterization tractable, we group all the non-volatile oxides ($\mathrm{Na_2O}$, $\mathrm{CaO}$, $\mathrm{SiO_2}$, etc.) as a single chemical component. This effective component is designated as rock and resides only in solid phase. Together with the volatile components H$_2$O and CO$_2$, the system in consideration is thus a two-phase system of three chemical components.

Directly parameterizing the full three-component system lacks a sound thermodynamic premise to start with (see Appendix \ref{apx:A}) and can obscure our understanding of the basic behaviors in dehydration and decarbonation of slab lithologies. Therefore, our strategy comprises two steps: (i) parameterize the simpler subsystems rock+H$_2$O and rock+CO$_2$ separately; (ii) synthesize the two subsystems into the full system (i.e., rock+H$_2$O+CO$_2$) by introducing additional parameters that account for the non-ideal thermodynamic mixing of H$_2$O and CO$_2$. This procedure is applied sequentially to lithologies relevant for subducting slabs: sediments, mid-ocean ridge basalts (MORB), gabbros, and mantle peridotites (Table~\ref{tab: composition}). 

\begin{table} 
\caption{Bulk Compositions for Representative Subducting Lithologies (wt\%)}
\label{tab: composition}
\begin{ruledtabular}
\begin{tabular}{l | c c c c c c c c c c}

& $\mathrm{SiO_2}$  & $\mathrm{TiO_2}$ & $\mathrm{Al_2O_3}$ & $\mathrm{FeO^T}$\footnotemark[1] & $\mathrm{MgO}$ & $\mathrm{CaO}$ & $\mathrm{Na_2O}$ & $\mathrm{K_2O}$ & H$_2$O & CO$_2$ \\
\hline
MORB & 49.33  & 1.46 & 15.31 & 10.33 & 7.41 & 10.82 & 2.53 & 0.19 & 2.61 & 2.88 \\
Gabbro & 48.64  & 0.87 & 15.48 & 5.96 & 8.84 & 12.02 & 2.69 & 0.096 & 2.58 & 2.84 \\
Sediment & 58.57  & 0.62 & 11.91 & 5.21 & 2.48 & 5.95 & 2.43 & 2.04 & 7.29 & 3.01 \\
Peridotite & 44.90  & 0.20 & 4.44 & 8.03 & 37.17 & 3.54 & 0.36 & 0.029 & -- & -- \\

\end{tabular}
\end{ruledtabular}
\footnotetext[1]{total iron including ferric and ferrous forms}
\end{table}

The subsystem parameterization employs ideal solution theory \citep[e.g.][]{Denbigh:1981aa, Rudge:2011aa, Keller:2016aa}. In an ideal solution, the dissolution of one chemical component does not affect the chemical potential of other components. Hence the partition coefficient ($K$) of component $i$ between solid and liquid phases can be represented as:
\begin{equation} \label{eq:sub-partition}
K^i = \frac{c_s^i}{c_l^i} = {c_{sat}^i(P)}\exp{\left[L_R(P)\left(\frac{1}{T} - \frac{1}{T_d^i(P)}\right)\right]},
\end{equation}
where $i$ represents either H$_2$O or CO$_2$. Notations of symbols are listed in Table~\ref{tab: symbol} and an exposition of the thermodynamic basis for equation~\eqref{eq:sub-partition} is provided in Appendix~\ref{apx:A}. $T_d^i(P)$ represents the onset temperature of effectively averaged dehydration or decarbonation in respective subsystems, and $L_R(P)$ is the corresponding enthalpy change divided by the gas constant ($R$) for the average reaction, hence termed the ``effective" enthalpy change. As a result, this functional form allows a focus on the leading-order behaviors of devolatilization by smoothing out detailed steps of mineral breakdown. Devolatilization is represented by a continuous process where volatiles dissolved in solid rocks exolve into co-existing liquid phases. We describe below how to extract the volatile saturation content ($c^i_{sat}$), onset temperature ($T^i_d$) and effective enthalpy change ($L_R$) from results of PerpleX.

Firstly, we use PerpleX \citep{Connolly:2002aa} to calculate pseudosection diagrams for the subsystems: rock+H$_2$O or rock+CO$_2$. For example, non-volatile oxide compositions (Table \ref{tab: composition}) from representative gabbros \citep{Hacker:2008aa} are adopted in our PerpleX calculations for the H$_2$O- and CO$_2$-saturated subsystems for the gabbroic lithology. Secondly, at each specific pressure, the maximum H$_2$O (or CO$_2$) content in gabbros can be extracted from the subsystem output from PerpleX as the corresponding volatile saturation ($c^i_{sat}$) content. Over the pressure range of interest (0.5--6~GPa) and with a pressure interval of $\sim$0.007 GPa, extraction at a series of specific pressure values yields a series of saturation values for H$_2$O (or CO$_2$), which can be regressed as a function of pressure. Thirdly, to further extract the values for $L_R$ and $T^i_d$ at that specific pressure, equation~\eqref{eq:sub-partition} is transformed to:
\begin{equation}
\ln K^i = \ln{{c_{sat}^i(P)}} + L_R(P)\left(\frac{1}{T} - \frac{1}{T_d^i(P)}\right).
\label{eq:sub-partition-log}
\end{equation}
According to equation~\eqref{eq:sub-partition-log}, when pressure is fixed at a specific value, $L_R$ is the slope and $-L_R/T^i_d + \ln{{c_{sat}^i}}$ is the intercept if linear regression is performed between $\ln K^i$ and $1/T$, where the $K^i$ values at different temperatures ($T$) come from the PerpleX results. Fourthly, with the known $c^i_{sat}$, $L_R$, and the intercept from above, $T_d^i$ can be obtained at the specific pressure. Similar to step two, the $T^i_d$ and linearly regressed $L_R$ values at various pressures are then fitted with polynomial functions of pressure over the same pressure range of interest. 

\begin{table}
\caption{Notations of Symbols in the Equations}
\label{tab: symbol}
\centering
\begin{ruledtabular}
\begin{tabular}{l l l}

\textbf{Symbol}  & \textbf{Meaning} & \textbf{Value and/or Unit} \\
\hline
   $c^i_s$  & mass fraction of volatile $i$ in solid phase \\
   $c^i_l$  & mass fraction of volatile $i$ in liquid  phase \\
   $c^i_{blk}$  & mass fraction of volatile $i$ in bulk system  \\
   $c^i_{sat}$  & saturated mass fraction of volatile $i$ in solid  \\
   $f$  & mass fraction of liquid phase in two-phase system \\
   $i$  & H$_2$O or CO$_2$   \\
   $K^i$  & ideal partition coefficient of volatile $i$  \\
   $\mathcal{K}^i$  & non-ideal partition coefficient of volatile $i$  \\
   $L_R$  & effective enthalpy change for devolatilization reactions & $\mathrm{K}$  \\  
   $P$  & pressure & GPa  \\ 
   $R$  & gas constant & 8.314 $\mathrm{J}\ \mathrm{K}^{-1} \ \mathrm{mol}^{-1}$  \\
   $T$  & temperature & $\mathrm{K}$ \\
   $T^i_d$  & onset temperature of devolatilization for volatile $i$ in subsystems & $\mathrm{K}$  \\      
   $\mathcal{T}^i_d$  & onset temperature of devolatilization for volatile $i$ in full systems & $\mathrm{K}$  \\  
   $\Delta T$  & temperature difference between decarbonation and dehydration & $\mathrm{K}$  \\ 
   $W_i$  & analogous Margules coefficients for volatile species $i$ & $\mathrm{J}\ \mathrm{mol}^{-1}$  \\  
   $\gamma^i$  & analogous activity coefficients for volatile species $i$  \\  

\end{tabular}
\end{ruledtabular}
\end{table}

The rock+H$_2$O and rock+CO$_2$ subsystems separately acquired through equation~\eqref{eq:sub-partition} and \eqref{eq:sub-partition-log} are then combined for the full-system parameterization. Mass balance for H$_2$O and CO$_2$ in full systems are:
\begin{equation} \label{eq: bulk1}
 c_{blk}^{\mathrm{H_2O}} = f c_l^{\mathrm{H_2O}} + (1 - f) c_l^{\mathrm{H_2O}} K^{\mathrm{H_2O}}, 
\end{equation}
\begin{equation} \label{eq: bulk2}
 c_{blk}^{\mathrm{CO_2}} = f (1-c_l^{\mathrm{H_2O}}) + (1 - f) (1 - c_l^{\mathrm{H_2O}} ) K^{\mathrm{CO_2}}, 
\end{equation}
where $c_{blk}^i$ is the bulk volatile content in the full system consisting of rock+H$_2$O+CO$_2$, and $f$ is the mass fraction of liquid phase consisting of only H$_2$O and CO$_2$. Note that because it is assumed that the volatile-free rock component resides only in the solid phase, a unity sum leads to $c_l^{\mathrm{CO_2}} = 1 - c_l^{\mathrm{H_2O}}$.  For any temperature, pressure, and bulk compositions ($c_{blk}^{\mathrm{H_2O}}$, $c_{blk}^{\mathrm{CO_2}}$), substituting the parameterized partition coefficients (equation~\eqref{eq:sub-partition}) makes equations~\eqref{eq: bulk1}--\eqref{eq: bulk2} closed, that is, only two unknowns ($f$ and $c_l^{\mathrm{H_2O}}$) remain.

The parameterization up to now assumes ideal solution behavior and does not involve nonideality. Adopting the representative bulk H$_2$O and CO$_2$ content for each lithology from Table~\ref{tab: composition} and solving equation~\eqref{eq: bulk1}--\eqref{eq: bulk2} over a $P$--$T$ range, we can calculate a pseudosection diagram describing $f$, $c^i_l$ and $c^i_s$ for each lithology. Parallel to this, adopting the same bulk compositions and employing PerpleX, we can compute pseudosection diagrams of the same type. Comparison between the results from parameterization and from PerpleX reveals discrepancies that are mainly due to non-ideal mixing. To make our parameterization better match PerpleX, two modifications are made to the subsystem partition coefficients ($K^i$) before they are used in the full system calculations in equations~\eqref{eq: bulk1} and \eqref{eq: bulk2}. The first modification is based on the requirement that when $c_{blk}^{\mathrm{H_2O}}=0$, the onset temperature of devolatilization for full system becomes that of decarbonation for the CO$_2$-only subsystem, whereas when $c_{blk}^{\mathrm{CO_2}}=0$, the full system similarly degenerates to H$_2$O-only subsystem. As such, the onset temperatures of dehydration ($T_d^{\mathrm{H_2O}}(P)$) and decarbonation ($T_d^{\mathrm{CO_2}}(P)$) are modified according to:
\begin{align}
\Delta T(P) &= T_d^{\mathrm{CO_2}}(P) - T_d^{\mathrm{H_2O}}(P),  \label{eqn:deltaT} \\
\mathcal{T}_{d}^{\mathrm{CO_2}}(P) &= T_d^{\mathrm{CO_2}}(P) - \Delta T(P) \left(\frac{c_{blk}^{\mathrm{H_2O}}}{c_{blk}^{\mathrm{H_2O}} + c_{blk}^{\mathrm{CO_2}}} \right)^2 \label{eqn:Tco2}, \\
\mathcal{T}_{d}^{\mathrm{H_2O}}(P) &= T_d^{\mathrm{H_2O}} (P)+ \Delta T(P) \left(\frac{c_{blk}^{\mathrm{CO_2}}}{c_{blk}^{\mathrm{H_2O}} + c_{blk}^{\mathrm{CO_2}}} \right)^2 \label{eqn:Th2o}, 
\end{align}
and $\mathcal{T}_{d}^{\mathrm{H_2O}}(P)$ and $\mathcal{T}_{d}^{\mathrm{CO_2}}(P)$ are substituted into equation~\eqref{eq:sub-partition} when used in full system calculations.

The second modification introduced is to account for the non-ideal mixing of H$_2$O and CO$_2$ in the liquid phase; readers are referred to Appendix~\ref{apx:A} for details. In doing so, the formalism of regular mixing for a Margules activity model is adopted \citep[e.g.][]{powell74}: 
\begin{equation} \label{eqn:non-ideal-partition}
\mathcal{K}^i = \frac{c_s^i}{c_l^i} = \gamma^i {c_{sat}^i(P)}\exp{\left[L_R(P)\left(\frac{1}{T} - \frac{1}{\mathcal{T}_d^i(P)}\right)\right]}, 
\end{equation}
\begin{equation} \label{eqn:h2o-activity}
R T \ln \gamma^{\mathrm{H_2O}} = W_{\mathrm{H_2O}}(P)\left(c_l^{\mathrm{CO_2}}\right)^2, 
\end{equation}
\begin{equation} \label{eqn:co2-activity}
R T \ln \gamma^{\mathrm{CO_2}} = W_{\mathrm{CO_2}}(P)\left(c_l^{\mathrm{H_2O}}\right)^2.
\end{equation}
Note that $W_{\mathrm{H_2O}}$ equals $W_{\mathrm{CO_2}}$ in the canonical regular mixing model, but we don't make this assumption here. Therefore, equations~\eqref{eqn:h2o-activity} and \eqref{eqn:co2-activity} adopt only the formalism of the regular mixing model, such that $W_{\mathrm{H_2O}}$ and $W_{\mathrm{CO_2}}$ are only analogous to the Margules coefficients. Likewise, the $\gamma^{\mathrm{H_2O}}$ and $\gamma^{\mathrm{CO_2}}$ are the analogous activity coefficients that account for the mismatch between results from the parameterization and from the full system PerpleX calculation. The mismatches are quantified according to equation~\eqref{eqn:non-ideal-partition} by dividing $\mathcal{K}^i$ derived from PerpleX by $K^i$ from our parameterization with $\mathcal{T}^i_d$ substituted (eq.~\eqref{eqn:Tco2}--\eqref{eqn:Th2o}). The values of $\gamma^i$ obtained in this way are then used to calculate $W_{\mathrm{H_2O}}$ and $W_{\mathrm{CO_2}}$ according to equations~\eqref{eqn:h2o-activity} and \eqref{eqn:co2-activity}. The $W_{\mathrm{H_2O}}$ and $W_{\mathrm{CO_2}}$ values are subsequently fitted as polynomial functions of pressure. Substituting the parameterization of $W_{\mathrm{H_2O}}$ and $W_{\mathrm{CO_2}}$ into equation~\eqref{eqn:non-ideal-partition}, we achieve an improved parameterization of partition coefficient that considers the effects of non-ideal mixing of H$_2$O and CO$_2$. The mass conservation equations for full systems become:
\begin{equation} \label{eq: gov1}
 c_{blk}^{\mathrm{H_2O}} = f c_l^{\mathrm{H_2O}} + (1 - f) c_l^{\mathrm{H_2O}} \mathcal{K}^{\mathrm{H_2O}}, 
\end{equation}
\begin{equation} \label{eq: gov2}
 c_{blk}^{\mathrm{CO_2}} = f (1-c_l^{\mathrm{H_2O}}) + (1 - f) (1 - c_l^{\mathrm{H_2O}} ) \mathcal{K}^{\mathrm{CO_2}}, 
\end{equation}
where the full system nonideal $\mathcal{K}^i$ replaces the subsystem ideal $K^i$, and the unknowns are only $c_l^{\mathrm{H_2O}}$ and $f$ as before.

In the next section, the above procedure is applied individually to each typical subducting lithologies (Table \ref{tab: composition}), whereby a parameterized thermodynamic module for subduction devolatilization is accomplished and can be readily interfaced with reactive flow modelling in our companion paper Part II. 

\section{Results} \label{sec:results}
\subsection{Gabbro} \label{sec:gab}
\subsubsection{Gabbro--H$_2$O Subsystem}
\begin{figure}[h!]
\centering
\includegraphics[width=0.8\columnwidth, keepaspectratio]{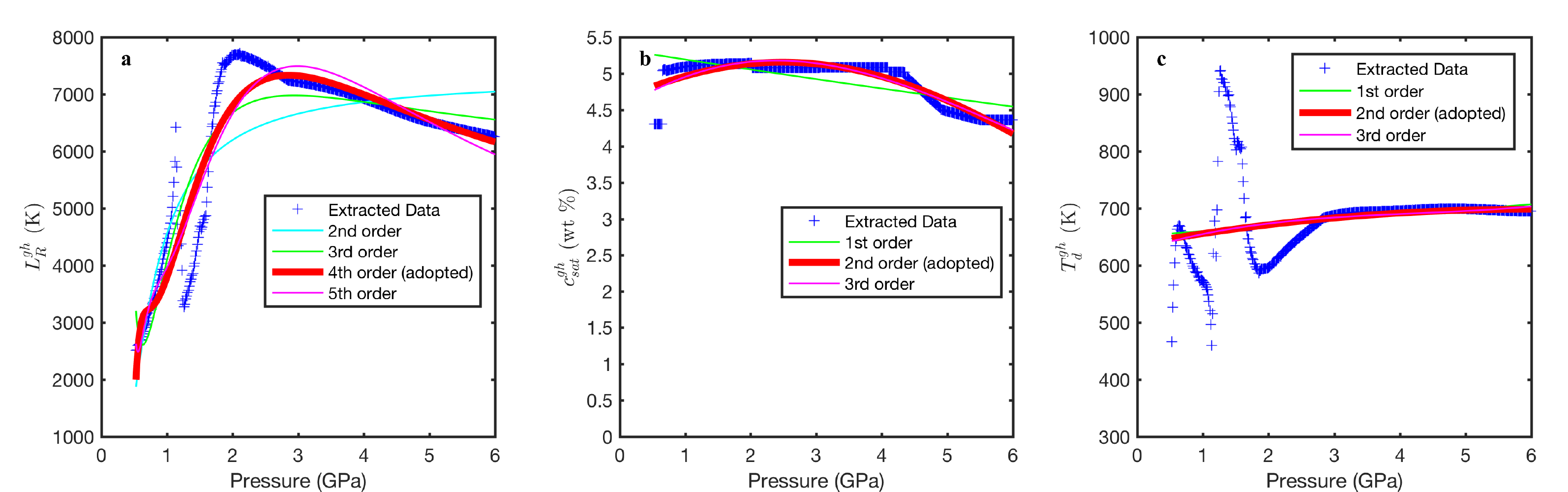} 
\caption{An example of fitting $L_R^{gh}$, $c_{sat}^{gh}$, and $T_d^{gh}$ of the gabbro--H$_2$O subsystem as polynomial functions of $1/P$ (a) and $P$ (b and c). The data used come from the PerpleX calculation of H$_2$O-saturated gabbro with oxide composition in Table \ref{tab: composition}. The thickened solid red lines denotes the polynomials adopted in this example case. Note that the fitting does not attempt to capture the wiggles of data around $\sim$1.2 GPa (a and c) because it will require extremely high-order polynomials and doesn't gain much. }
\label{fig: regression-ex}
\end{figure}

\begin{figure}[h!]
\centering
\includegraphics[width=0.65\columnwidth, keepaspectratio]{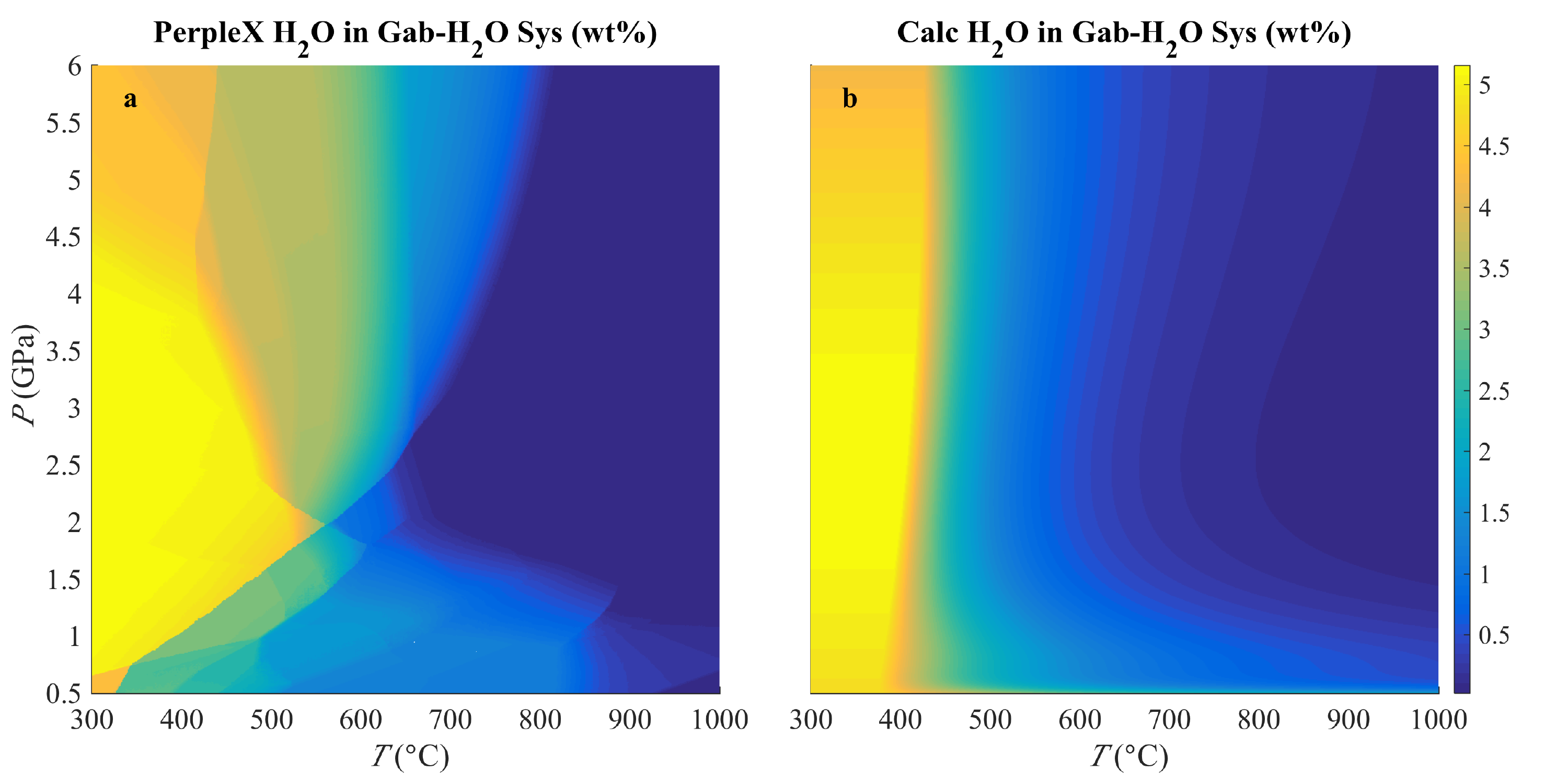} 
\caption{Comparison between the gabbro H$_2$O contents ($c_s^{\mathrm{H_2O}}$) derived from PerpleX (a) and our parameterization (b) for the gabbro--H$_2$O subsystem. ``Gab" in the titles is short for gabbro, and the same as below. }
\label{fig: gab-h2o}
\end{figure}

The representative, volatile-free, bulk composition for gabbro is taken from \cite{Hacker:2008aa} and a pseudosection is calculated by PerpleX assuming H$_2$O saturation (Fig. \ref{fig: gab-h2o}a). Following the strategy in Section \ref{sec:strategy}, we can extract from the PerpleX result the values of $c_{sat}^{\mathrm{H_2O}}$, $T_d^{\mathrm{H_2O}}$ and $L_R$ for each incremented pressure ($P$) in the pressure range of interest (0.5--6 GPa). In fitting these values as polynomial functions of pressure ($P$), higher-order polynomials fit better, but also come with increased likelihood of yielding bad extrapolations. Thus, a balance needs to be maintained between good accuracy of fitting and low orders of polynomials. For all the regressions in this paper, the ``best fit" is chosen by inspection when it achieves such a balance. Since $L_R$ controls the rate of dehydration, we fit it as a polynomial of $1/P$ rather than $P$, such that the parameterized $L_R$ leads to neither extremely sharp dehydration nor non-dehydration when extrapolated to high pressures. Figure \ref{fig: regression-ex} shows an example of how $L_R$ of the gabbro--H$_2$O subsystem is regressed as a function of $1/P$. With the data of $L_R$ extracted from PerpleX, we experiment with $1/P$ polynomials from low to high orders. As shown in Figure \ref{fig: regression-ex}, the fourth order polynomial is adopted in this case: 
\begin{equation} \label{eq:lr-gh}
\ln(L_R^{gh}(P)) = b_0 /P^4 + b_1 /P^3 + b_2 /P^2 + b_3 /P + b_4,
\end{equation}
and $c_{sat}^{\mathrm{H_2O}}$ and $T_d^{\mathrm{H_2O}}$ are regressed in a similar way:
\begin{equation} \label{eq:sat-gh}
\ln(c_{sat}^{gh}(P)) = a_0 P^2 + a_1 P + a_2, 
\end{equation}
\begin{equation} \label{eq:td-gh}
T_d^{gh}(P) = c_0 P^2 + c_1 P + c_2,
\end{equation}
where the superscript ``\textit{gh}" denotes ``gabbro--H$_2$O", pressures ($P$) are in GPa, temperatures ($T$) in $K$, and all the regressed coefficients are given in Table \ref{tab:gabbro}.

By substituting equations~\eqref{eq:lr-gh}--\eqref{eq:td-gh} into \eqref{eq:sub-partition} and noting that $c_l^{\mathrm{H_2O}}=1$ in the subsystem, the H$_2$O content in gabbro ($c_s^{\mathrm{H_2O}}$) is computed and plotted in Figure \ref{fig: gab-h2o}b for comparison with the PerpleX result in Figure~\ref{fig: gab-h2o}a. It can be seen that the phase diagram from PerpleX contains multiple dehydration reactions diagnostic of both continuous and discontinuous reduction in the H$_2$O content of gabbro \citep{Schmidt:2014aa}, whereas the parameterization is an average of these reactions. The effectively averaged reaction is characterized by an onset dehydration temperature ($T_d^{\mathrm{H_2O}}$), and a gradual H$_2$O content reduction the rate of which is controlled by $L_R$. Therefore, the parameterized $T_d^{\mathrm{H_2O}}$, $c_{sat}^{\mathrm{H_2O}}$, and $L_R$ capture the gross behavior of the subsystem with respect to initial dehydration, maximum H$_2$O content, and the smoothness of dehydration reaction. Experimental studies on MORB \citep{Schmidt:1998aa}, which is compositionally similar to gabbro, show that below $\sim$2.2--2.4 GPa where amphiboles break down, there is a larger number of dehydration reactions than above $\sim$2.4 GPa, and this gives rise to a gradual reduction in gabbro H$_2$O content. Above this pressure, the varieties of hydrous phases reduce to dominantly lawsonites and phengites, leading to less dehydration reactions. Since phengites decomposition occurs at higher temperatures than lawsonites and principally participate in fluid-absent melting, the H$_2$O release above $\sim$2.2--2.4 GPa is mostly caused by lawsonite breakdown between $\sim$700--800 $^{\circ}$C. The reduction of gabbro H$_2$O content accordingly becomes sharper in this pressure range. These two features of dehydration under low and high pressure conditions are captured in Figure \ref{fig: gab-h2o}b of our parameterization.

\subsubsection{Gabbro--CO$_2$ Subsystem} \label{sec:gab-co2}
\begin{figure}[h!]
\centering
\includegraphics[width=0.65\columnwidth, keepaspectratio]{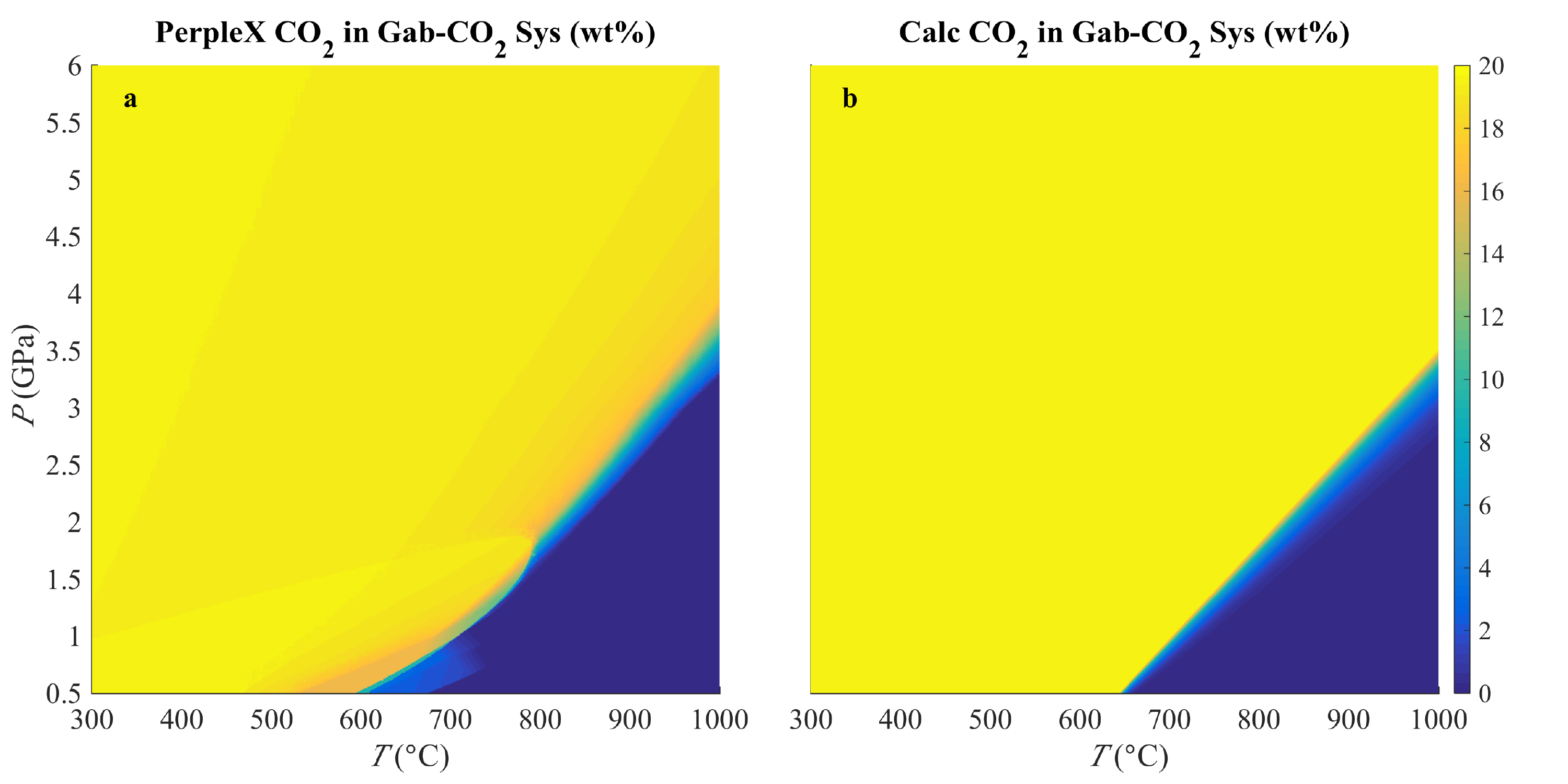} 
\caption{Comparison between the gabbro CO$_2$ contents ($c_s^{\mathrm{CO_2}}$) derived from PerpleX (a) and our parameterization (b) for the gabbro--CO$_2$ subsystem.}
\label{fig: gab-co2}
\end{figure}

The same volatile-free bulk composition as in the gabbro--H$_2$O subsystem is used in the parameterization of the gabbro--CO$_2$ subsystem. Figure \ref{fig: gab-co2}a illustrates the result of gabbro CO$_2$ content calculated by PerpleX under CO$_2$ saturation conditions. The same approach as above is used to extract $c_{sat}^{\mathrm{CO_2}}$, $L_R$, and $T_d^{\mathrm{CO_2}}$, and the following polynomials best fit the data extracted from PerpleX:
\begin{equation} \label{eqn:gab-co2-sat}
c_{sat}^{gc}(P) = a_0, 
\end{equation}
\begin{equation}
\ln(L_R^{gc}(P)) = b_0 /P + b_1, 
\end{equation}
\begin{equation}
T_d^{gc}(P) = c_0 P + c_1,
\end{equation}
where the superscript ``$gc$'' represents ``gabbro--CO$_2$''. Because the varieties of CO$_2$-bearing minerals are far less than those of hydrous minerals, the variation of saturated CO$_2$ content in the subsystem is minimal (Fig. \ref{fig: gab-co2}a). In fact, the CO$_2$-containing minerals in Figure \ref{fig: gab-co2}a are only aragonite, dolomite, and magnesite. Under CO$_2$ saturation conditions, it is thus the Mg and Ca content that determines the rock CO$_2$ content. Given that the Mg and Ca mass fractions in the bulk lithology are assumed to be constant, the $c_{sat}^{\mathrm{CO_2}}$ values are virtually unchanging. Vanishingly small variations of $c_{sat}^{\mathrm{CO_2}}$ occur in the PerpleX plot (Fig. \ref{fig: gab-co2}a) due to minor variations of total amount of Mg and Ca (and sometimes Fe) that participates in carbonate formation. At a fixed pressure, the stable carbonate phase(s) change from aragonite to dolomite + magnesite, and to dolomite with increasing temperature. However, these mineralogical transformations involve mostly the adjustment of Ca and Mg proportions in the carbonates (solid solutions), rather than the total amount of Ca and Mg that forms carbonates. In consequence, the CO$_2$ content stays more or less constant until a temperature is reached where all the CO$_2$-bearing minerals become unstable, that is, the breakdown of dolomite. Beyond this temperature, no minerals can hold CO$_2$ and its content in gabbro sharply reduces to almost zero, as illustrated in Figure \ref{fig: gab-co2}a. Our parameterization in Figure \ref{fig: gab-co2}b captures this decarbonation feature, with a constant $c_{sat}^{\mathrm{CO_2}}$, a decarbonation curve $T_d^{\mathrm{CO_2}}(P)$ approximating dolomite breakdown, and large $L_R^{gc}(P)$ accounting for the sharp reduction of CO$_2$ beyond $T_d^{\mathrm{CO_2}}$. The regressed parameters are given in Table \ref{tab:gabbro}. 

\subsubsection{Gabbro--H$_2$O--CO$_2$ Full System} \label{sec:gab-co2-h2o}
\begin{figure}[h!]
\centering
\includegraphics[width=0.65\columnwidth, keepaspectratio]{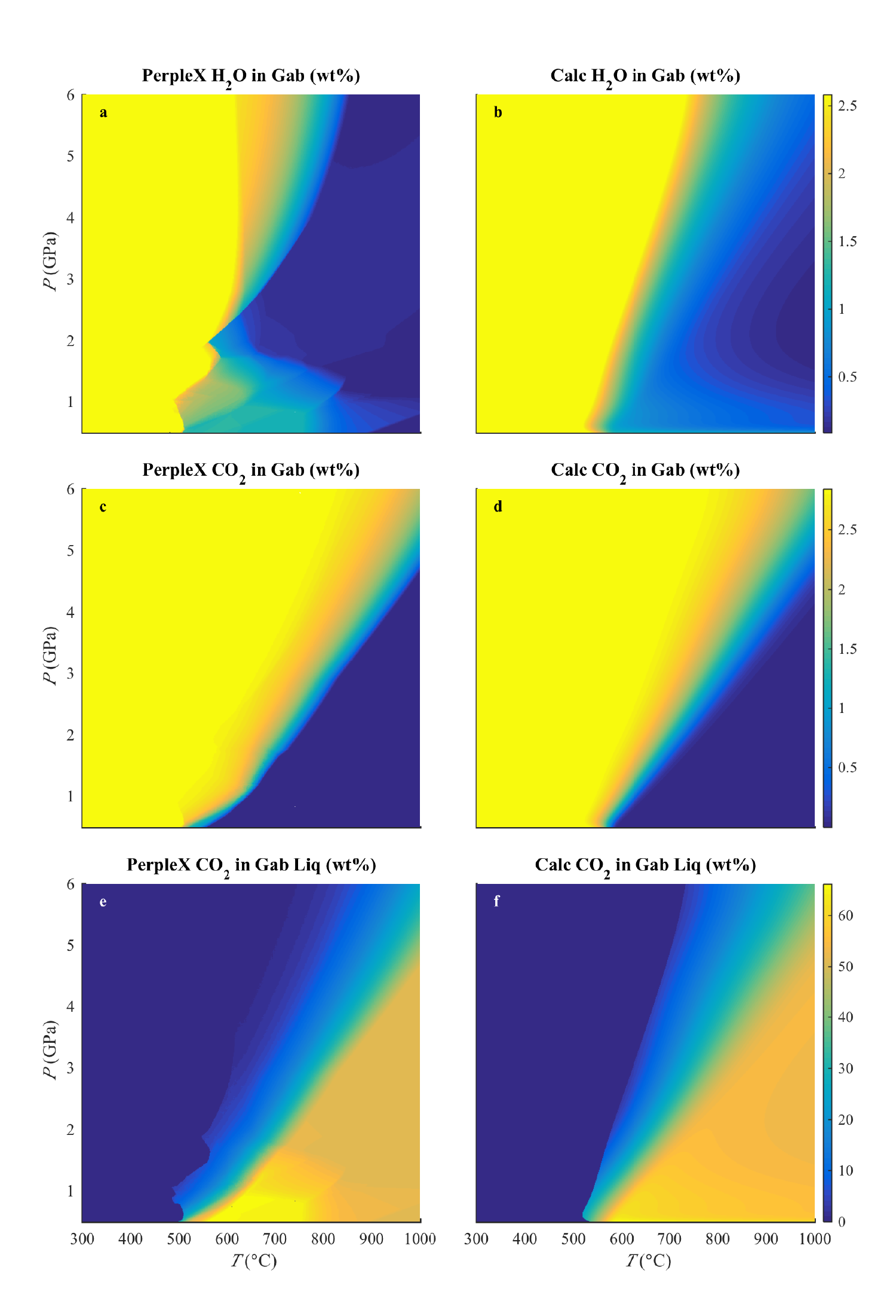} 
\caption{Comparison between the results from PerpleX (left panels) and our parameterization (right panels) for the gabbro--H$_2$O--CO$_2$ full system. Panels (a)--(b) are for H$_2$O content in solid rocks ($c_s^{\mathrm{H_2O}}$), (c)--(d) are for CO$_2$ content in solid rocks ($c_s^{\mathrm{CO_2}}$), and (e)--(f) are for CO$_2$ content in the liquid volatile phase ($c_l^{\mathrm{CO_2}}$). Note that $c_l^{\mathrm{H_2O}} = 1- c_l^{\mathrm{CO_2}}$, so the comparability of H$_2$O content in the liquid phase can be deduced from that of CO$_2$.}
\label{fig: gab-h2o-co2}
\end{figure}

With the parameters acquired above, the subsystem partition coefficients ($K^i$) can be calculated according to equation~\eqref{eq:sub-partition}, and a tentative full system pseudosection at specified $c_{blk}^{\mathrm{H_2O}}$ and $c_{blk}^{\mathrm{CO_2}}$ (Table \ref{tab: composition}) can be computed according to equations~\eqref{eq: bulk1}--\eqref{eq: bulk2}. Using the same bulk composition as input to PerpleX independently yields another pseudosection that differs from the one calculated by parameterization. Following the strategy in Section \ref{sec:strategy}, the discrepancies are used to parameterize analogous $W_{\mathrm{H_2O}}$ and $W_{\mathrm{CO_2}}$ (eq.~\eqref{eqn:h2o-activity}--\eqref{eqn:co2-activity}) as polynomial functions of pressure. After experimenting with polynomials of increasing order, we find the following best fit the data: 
\begin{equation}
\ln(-W_{\mathrm{H_2O}}) = d_0 P^4 + d_1 P^3 + d_2 P^2 + d_3 P + d_4,
\end{equation}
\begin{equation}
\ln(-W_{\mathrm{CO_2}}) = e_0 P^4 + e_1 P^3 + e_2 P^2 + e_3 P + e_4.
\end{equation}
Table \ref{tab:gabbro} lists the relevant coefficients from data fitting. Figure \ref{fig: gab-h2o-co2} demonstrates that the fully parameterized thermodynamic calculation compares well with that from PerpleX. In particular, as shown in Figure \ref{fig: gab-h2o-co2}c \& e, pressure increase at a specific temperature stablizes carbonate minerals and makes coexisting liquid deficient in CO$_2$. Moreover, temperature increase at a fixed pressure leads to CO$_2$ enrichment in the equilibrated liquid phase. These two leading-order features during devolatilization are observed by \citet{Molina:2000aa} and reflected in our parameterization (Fig. \ref{fig: gab-h2o-co2}d \& f).

\begin{table}[h!]
\caption{Regression Results{\footnote{From left to right, the values correspond to polynomial coefficients in increasing order of subscript; same for other tables.}} for Gabbro--H$_2$O--CO$_2$ System}
\label{tab:gabbro}
\centering
\begin{ruledtabular}
\begin{tabular}{l c c c c c c c}

     & \multicolumn{5}{c@{\hspace{1cm}}|}{H$_2$O} & \multicolumn{2}{c}{CO$_2$} \\
\hline
   $c_{sat}$ & -- & -- & $-$0.0176673 & 0.0893044 & \multicolumn{1}{c@{\hspace{1cm}}|}{1.52732} & -- & 19.3795 \\
   $L_R$ & $-$1.81745 & 7.67198 & $-$10.8507 & 5.09329 & \multicolumn{1}{c@{\hspace{1cm}}|}{8.14519} & 0.661119 & 10.9216 \\
   $T_d$ & -- & -- & $-$1.72277 & 20.5898 & \multicolumn{1}{c@{\hspace{1cm}}|}{637.517} & 118.286 & 857.854 \\    
\hline
   $W_{\mathrm{H_2O}}$ & -- & -- & $-$0.03522 & 0.5204 & \multicolumn{1}{c@{\hspace{1cm}}}{$-$2.381} & 3.64 & $-$9.995 \\ 
   $W_{\mathrm{CO_2}}$ & -- & -- & 0.009474 & $-$0.1576 & \multicolumn{1}{c@{\hspace{1cm}}}{0.9418} & $-$2.283 & 13.37 \\       

\end{tabular}
\end{ruledtabular}
\end{table}

\subsection{Basalt Devolatilization}
The same procedure as for gabbro is followed in parameterizing the MORB--CO$_2$--H$_2$O system, utilizing the volatile-free bulk composition for MORB from Table \ref{tab: composition}. Since all the data fitting in this contribution is in polynomial form, to avoid repetition, we detail the functional forms of parameterization for lithologies other than gabbro in Appendix \ref{apx:B}, and focus on discussing the parameterized results in the succeeding text.

\begin{figure}[h!]
\centering
\includegraphics[width=0.65\columnwidth, keepaspectratio]{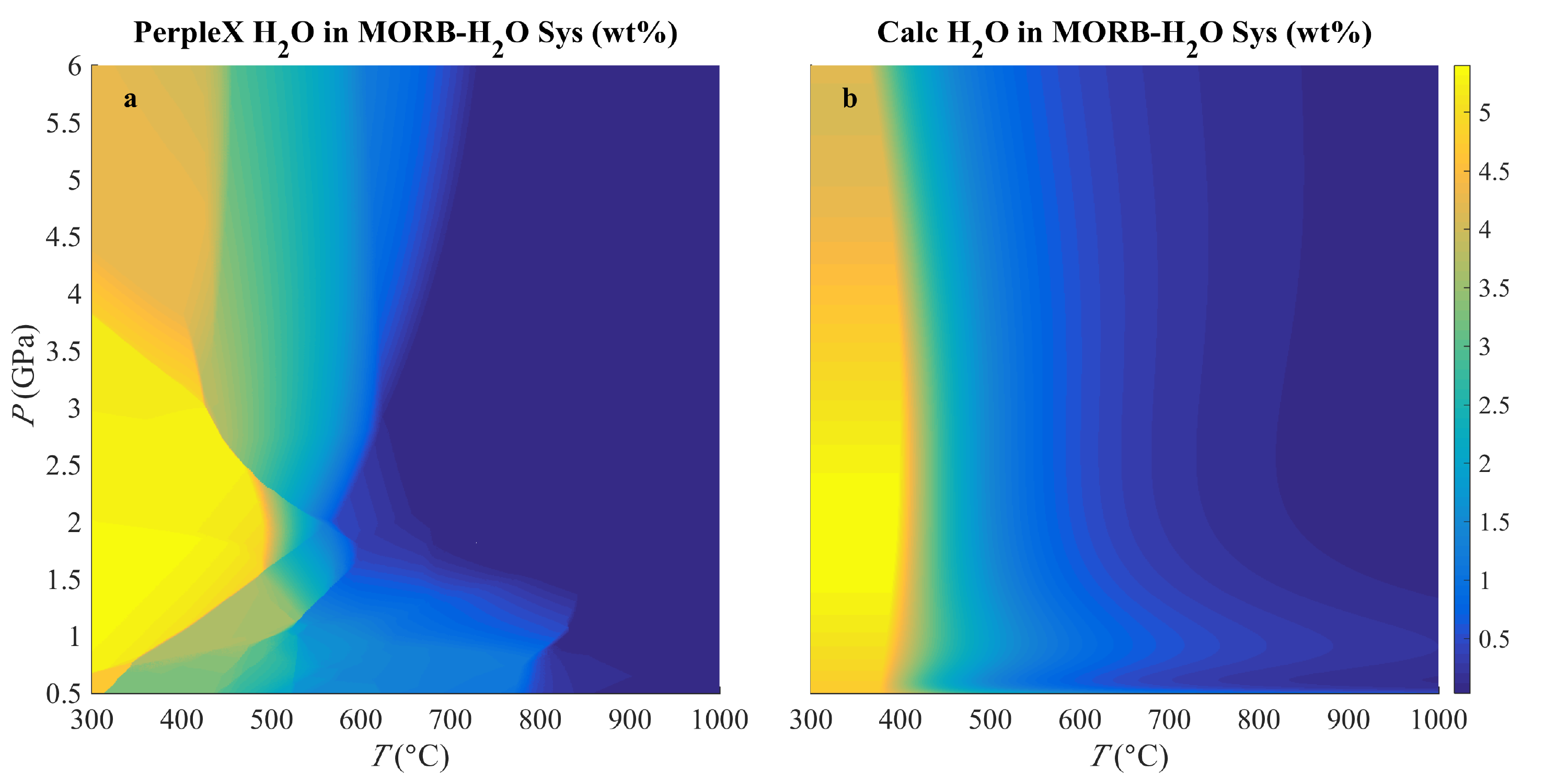} 
\caption{Comparison between the MORB H$_2$O contents ($c_s^{\mathrm{H_2O}}$) derived from PerpleX (a) and our parameterization (b) for the MORB--H$_2$O subsystem.}
\label{fig: morb-h2o}
\end{figure}

A comparison between the parameterized and PerpleX results for the MORB--H$_2$O subsystem is illustrated in Figure \ref{fig: morb-h2o}. Considering that basalts and gabbros are compositionally close to each other, the phase diagrams in Figures \ref{fig: gab-h2o}a and \ref{fig: morb-h2o}a are much alike, and the fitting polynomials and plot from parameterization are accordingly similar. Therefore, the experimental validation of parameterization for gabbro devolatlization in Section \ref{sec:gab} applies to the representative basalts in this section.

\begin{figure}[h!]
\centering
\includegraphics[width=0.65\columnwidth, keepaspectratio]{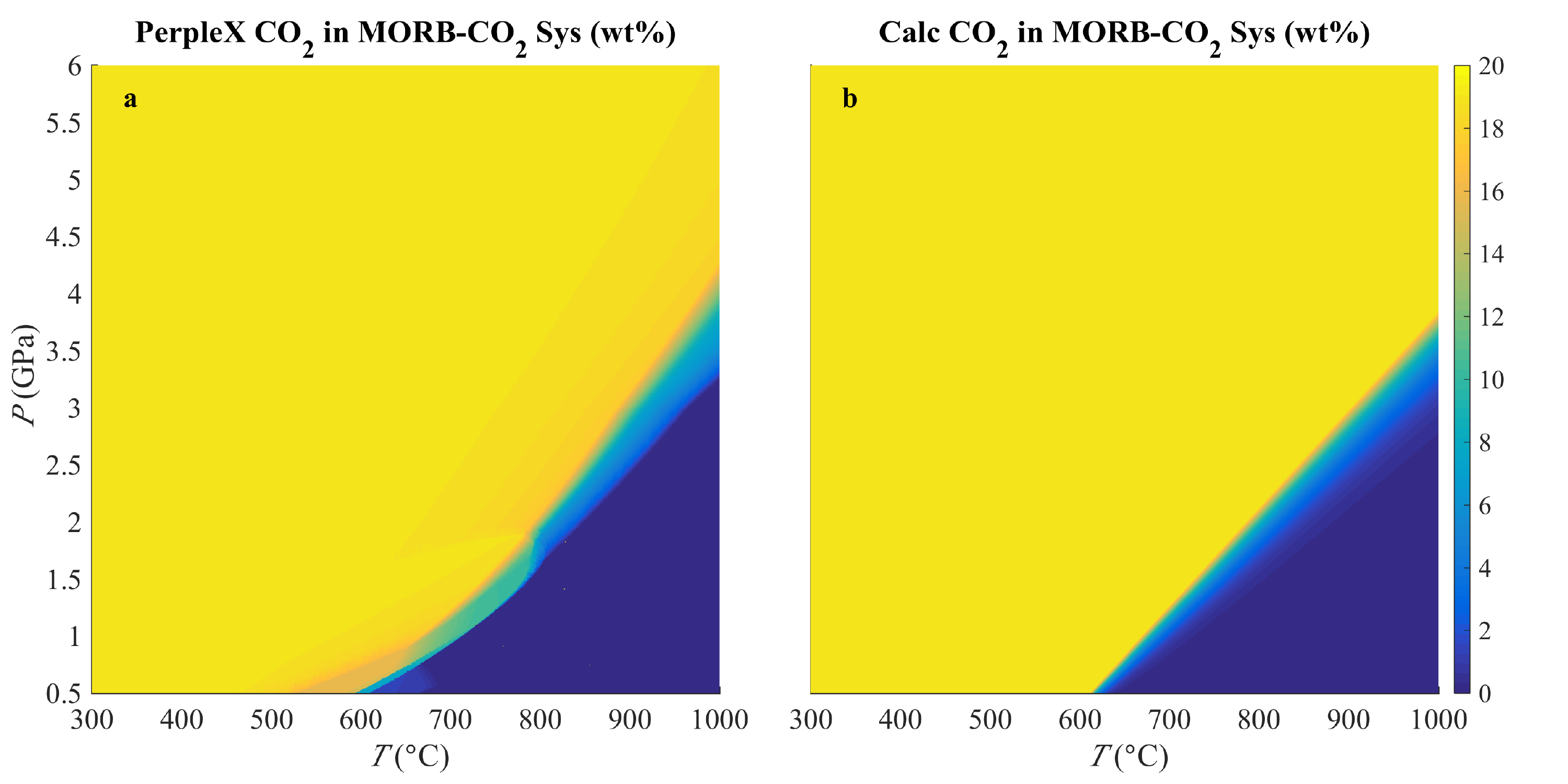} 
\caption{Comparison between the MORB CO$_2$ contents ($c_s^{\mathrm{CO_2}}$) derived from PerpleX (a) and our parameterization (b) for the MORB--CO$_2$ subsystem.}
\label{fig: morb-co2}
\end{figure}

For the CO$_2$-only subsystem, Figure \ref{fig: morb-co2} compares the parameterization and PerpleX results. As noted in the instance for gabbro--CO$_2$ subsystem, due to the limited number of the varieties of carbonate minerals, the saturated CO$_2$ content ($c_{sat}^{\mathrm{CO_2}}$) is approximately constant, along with large $L_R$ values describing the sharp breakdown of the last carbonate mineral---dolomite. 

In synthesizing the subsystem parameterizations above to the full MORB--H$_2$O--CO$_2$ system, the bulk MORB composition in Table \ref{tab: composition} is adopted in generating the PerpleX result which is further used when parameterizing the relevant $W_{\mathrm{H_2O}}$ and $W_{\mathrm{CO_2}}$. Polynomial forms for fitting the $W_{\mathrm{H_2O}}$ and $W_{\mathrm{CO_2}}$ data are provided in Appendix~\ref{apx:B}. Figure \ref{fig: morb-h2o-co2} compares the the results from PerpleX and the parameterization for the full system. In addition to the two leading-order features outlined in Section \ref{sec:gab-co2-h2o}, comparison between Figure \ref{fig: morb-h2o-co2}a and b shows that our parameterization favors H$_2$O release at low pressure and temperature conditions ($\sim$0.5--1.0 GPa and $\sim$500--700 $^{\circ}$C) relative to PerpleX. However, such a difference in low $P$ \& $T$ regime does not affect modelling subduction-zone devolatilization because the global geotherms for the slab MORB layer all lie above this $P$ \& $T$ region \citep[e.g., Figure 5 in][]{Keken:2011aa}.
 
\begin{figure}[h!]
\centering
\includegraphics[width=0.65\columnwidth, keepaspectratio]{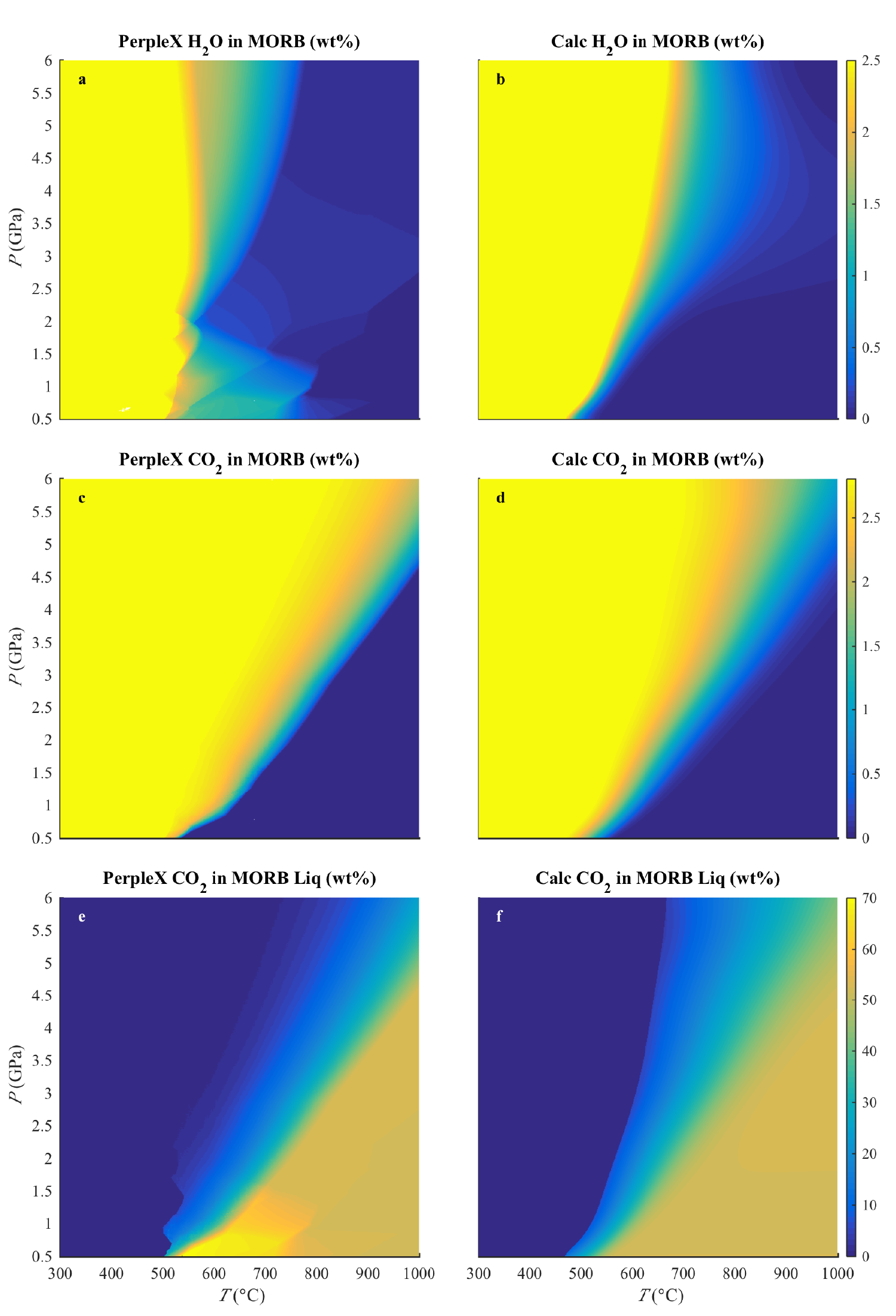} 
\caption{Comparison between the results from PerpleX (left panels) and our parameterization (right panels) for the MORB--H$_2$O--CO$_2$ full system. Panels (a)--(b) are for H$_2$O content in solid rocks ($c_s^{\mathrm{H_2O}}$), (c)--(d) are for CO$_2$ content in solid rocks ($c_s^{\mathrm{CO_2}}$), and (e)--(f) are for CO$_2$ content in the liquid volatile phase ($c_l^{\mathrm{CO_2}}$).}
\label{fig: morb-h2o-co2}
\end{figure}

\subsection{Sediment Devolatilization}
Although sedimentary layers are only a few 100 m thick atop subducting slabs, they contain high abundance of incompatible minor and trace elements that are crucial in characterizing arc lava genesis \citep{Schmidt:2014aa}. As such, most experimental studies focus on the melting behavior of subducted sediments, rather than devolatilization \citep[e.g.,][]{Mann:2015aa, Thomsen:2008aa, Tsuno:2012aa}. Our parameterization is thus compared below with previous modelling results and the limited experimental data available.

\begin{figure}[h!]
\centering
\includegraphics[width=0.65\columnwidth, keepaspectratio]{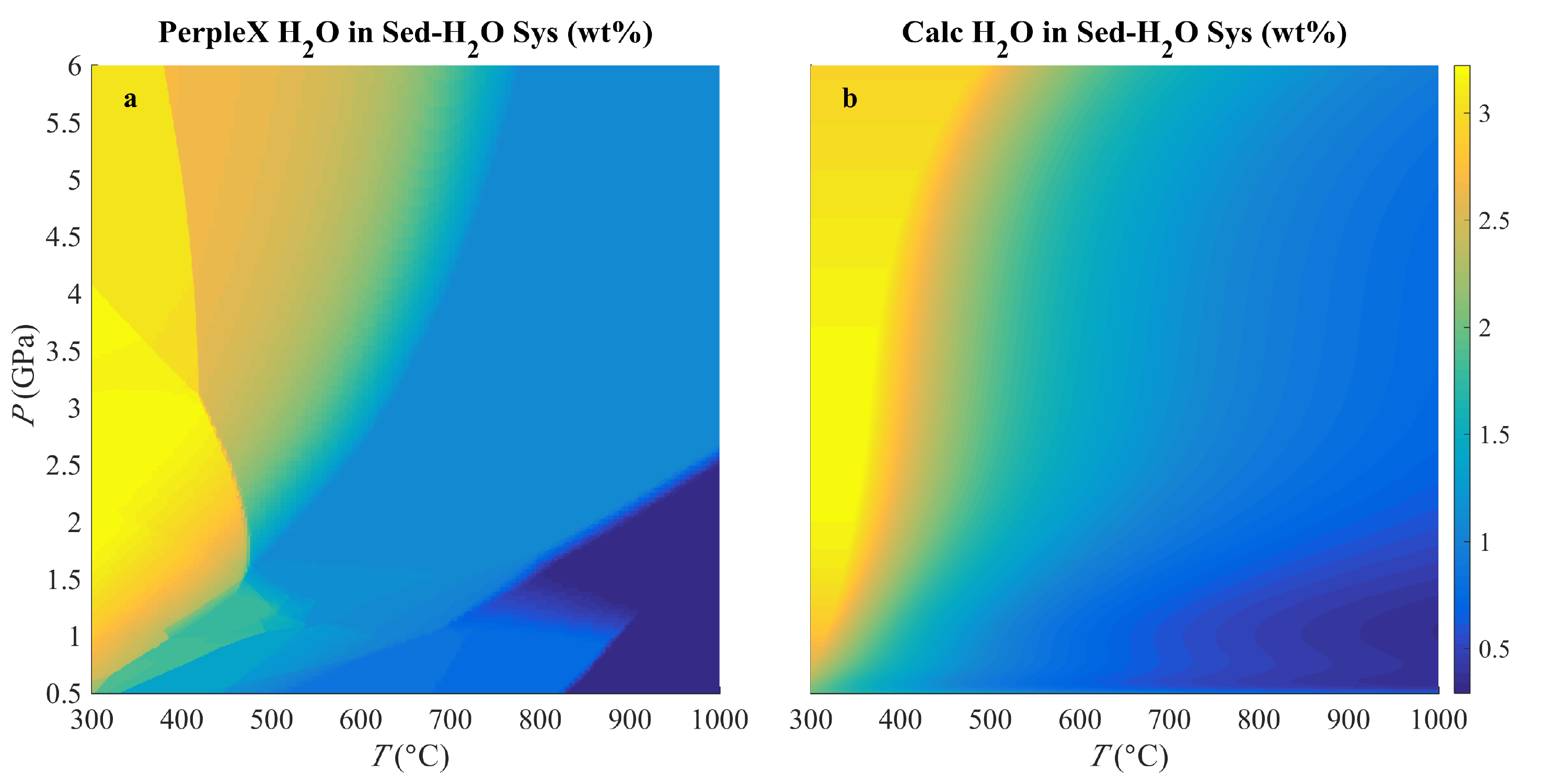} 
\caption{Comparison between the sediment H$_2$O contents ($c_s^{\mathrm{H_2O}}$) derived from PerpleX (a) and our parameterization (b) for the sediment--H$_2$O subsystem. ``Sed" in the titles is short for sediment, and the same as below. }
\label{fig: sed-h2o}
\end{figure}

Figure \ref{fig: sed-h2o} shows a comparison of the parameterized and PerpleX results for the sediment--H$_2$O subsystem. As hydrous mineral phases that can be stable in metapelites are abundant (e.g., micas, talc, chloritoid, chlorite, etc.), and they involve extensive solid solutions, the H$_2$O content changes smoothly. In addition, at pressures above $\sim$1.5 GPa, metapelite dehydration is dominated by the breakdown of two major hydrous phases---amphiboles and micas (Fig. \ref{fig: sed-h2o}a), so the parameterized initial dehydration curve $T_d^{sh}$ (see Appendix \ref{apx:B}) is approximately their average (Fig. \ref{fig: sed-h2o}b). Particularly, as \citet{Schmidt:2014aa} inferred from the experiments by \citet{Chmielowski:2010aa}, metapelite contains about 2.0 wt\% H$_2$O at 3 GPa and 600 $^{\circ}$C, which is consistent with our parameterization in Figure \ref{fig: sed-h2o}b. However, in contrast to the PerpleX calculation, our parameterization does not predict complete dehydration at high temperatures ($>$ $\sim$900 $^{\circ}$C) and lower pressures ($<$ $\sim$2 GPa). Nonetheless, this limitation is unconcerning for the current parameterization because partial melting is expected in this $P$\slash $T$ range. Since both the parameterization and PerpleX calculation ignore partial melting, it is thus not necessary to match them in the $P$\slash $T$ region where partial melting would occur. 

\begin{figure}[ht!]
\centering
\includegraphics[width=0.65\columnwidth, keepaspectratio]{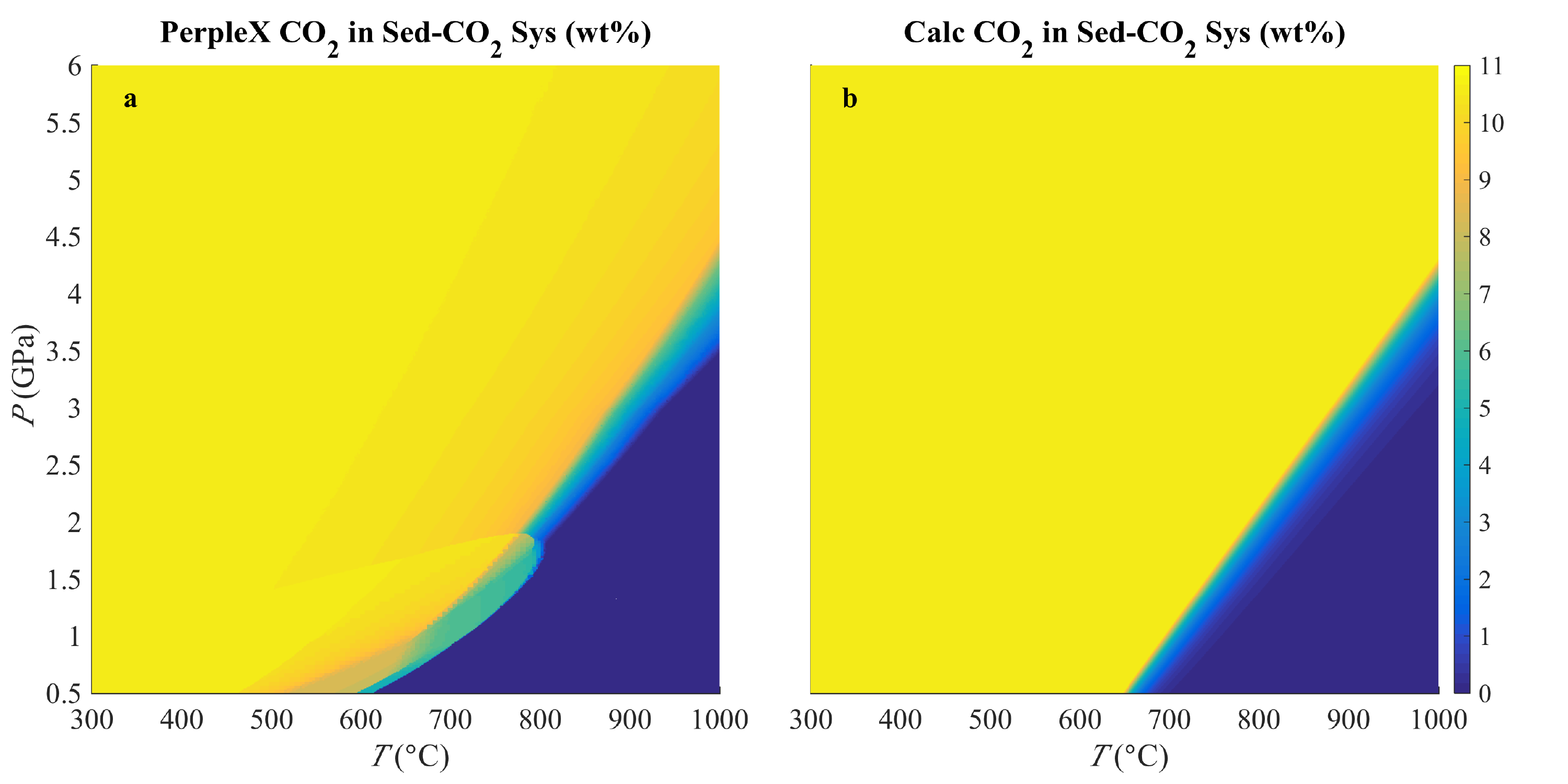} 
\caption{Comparison between the sediment CO$_2$ contents ($c_s^{\mathrm{CO_2}}$) derived from PerpleX (a) and our parameterization (b) for the sediment--CO$_2$ subsystem.}
\label{fig: sed-co2}
\end{figure}

For the sediment--CO$_2$ subsystem, as the major carbonate phases in metapelites are still calcite\slash aragonite, magnesite, and dolomite \citep{Schmidt:2014aa}, the formulae (see Appendix~\ref{apx:B}) used during parameterization are the same as those for the basalt--CO$_2$ subsystem. A comparison of the PerpleX and parameterized results is illustrated in Figure \ref{fig: sed-co2}, where the curve of the onset of decarbonation ($T_d^{sc}(P)$) approximates the breakdown of dolomite.

\begin{figure}[h!]
\centering
\includegraphics[width=0.65\columnwidth, keepaspectratio]{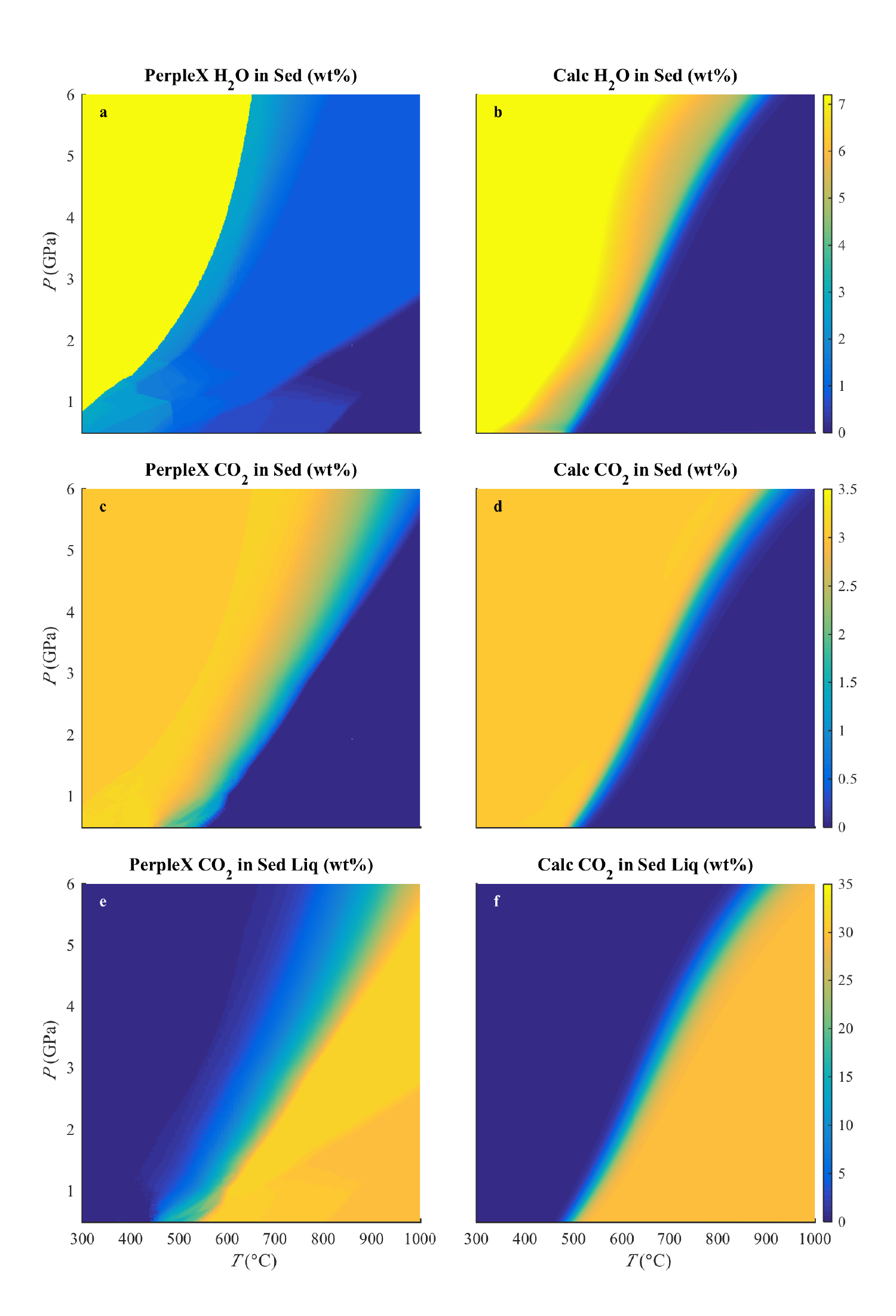} 
\caption{Comparison between the results from PerpleX (left panels) and our parameterization (right panels) for the sediment--H$_2$O--CO$_2$ full system. Panels (a)--(b) are for H$_2$O content in solid rocks ($c_s^{\mathrm{H_2O}}$), (c)--(d) are for CO$_2$ content in solid rocks ($c_s^{\mathrm{CO_2}}$), and (e)--(f) are for CO$_2$ content in the liquid volatile phase ($c_l^{\mathrm{CO_2}}$).}
\label{fig: sed-h2o-co2}
\end{figure}

Figure~\ref{fig: sed-h2o-co2} shows the parameterized results for the full sediment--H$_2$O--CO$_2$ system. Comparison of Figure \ref{fig: sed-h2o}b and \ref{fig: sed-h2o-co2}b suggests that addition of carbonates into the system facilitates complete H$_2$O release at high temperatures. Close inspection of the PerpleX result in Figure \ref{fig: sed-h2o-co2}a reveals that there are two major discontinuities in H$_2$O loss: one from H$_2$O content 7.2 wt\% to $\sim$1 wt\%, and the other from $\sim$1 wt\% to almost 0 wt\%. On the other hand, our parameterized result in Figure \ref{fig: sed-h2o-co2}b depicts a gradual reduction of H$_2$O content from 7.2 wt\% to almost 0 wt\%. In addition, the parameterized onset temperature of H$_2$O loss is slightly higher than that from PerpleX, but the temperature of complete H$_2$O loss is lower for the parameterization than for PerpleX, which is consistent with the formalism adopted to effectively average devolatilization reactions (Appendix \ref{apx:A}). Further, comparison between Figure \ref{fig: sed-co2}b and \ref{fig: sed-h2o-co2}d suggests that addition of H$_2$O into the CO$_2$-only subsystem reduces decarbonation temperatures and thus promotes CO$_2$ release, facilitating the scenario of infiltration-driven decarbonation \citep{Gorman:2006aa}. At 750--850 $^{\circ}$C and $\sim$2 GPa, experimental studies yield $X_{\mathrm{CO_2}}$ (mole fraction of CO$_2$ in liquid phase) estimates of 0.14 and 0.10--0.19 equilibrated with carbonated pelites and basalts respectively \citep{Molina:2000aa, Thomsen:2008ab}. Since $X_{\mathrm{CO_2}} \approx$ 0.14 converts to $\sim$28 wt\% of CO$_2$, our parameterization illustrated in Figure \ref{fig: sed-h2o-co2}f is compatible with these experimental constraints. Overall, with increasing temperature and decreasing pressure, Figure \ref{fig: sed-h2o-co2}f shows that the equilibrated liquid phase becomes more and more CO$_2$-rich \citep{Thomsen:2008ab}.

\subsection{Devolatilization of Peridotite in Upper Mantle}
Dehydration of hydrated slab mantle has been invoked as a mechanism for CO$_2$ release via infiltration-driven decarbonation \citep[e.g.,][]{Gorman:2006aa, Kerrick:2001ab}; moreover, the seismic activities within double seismic zones are also attributed to dehydration embrittlement in the slab upper mantle \citep[e.g.,][]{Peacock:2001aa}. Inferred from seismic data, the hydration state of the slab lithospheric mantle is highly variable and thus uncertain \citep[see][]{Garth:2017aa, Korenaga:2017aa}, and the carbonation state even more so \citep[e.g.,][]{Kerrick:1998aa}. Therefore, for the purpose of modelling slab devolatilization, the basal lithospheric mantle layer is treated as a water supplier that gives rise to H$_2$O infiltration into the overlying lithologies, and parameterization is performed on the H$_2$O-only subsystem. 

\begin{figure}[h!]
\centering
\includegraphics[width=0.65\columnwidth, keepaspectratio]{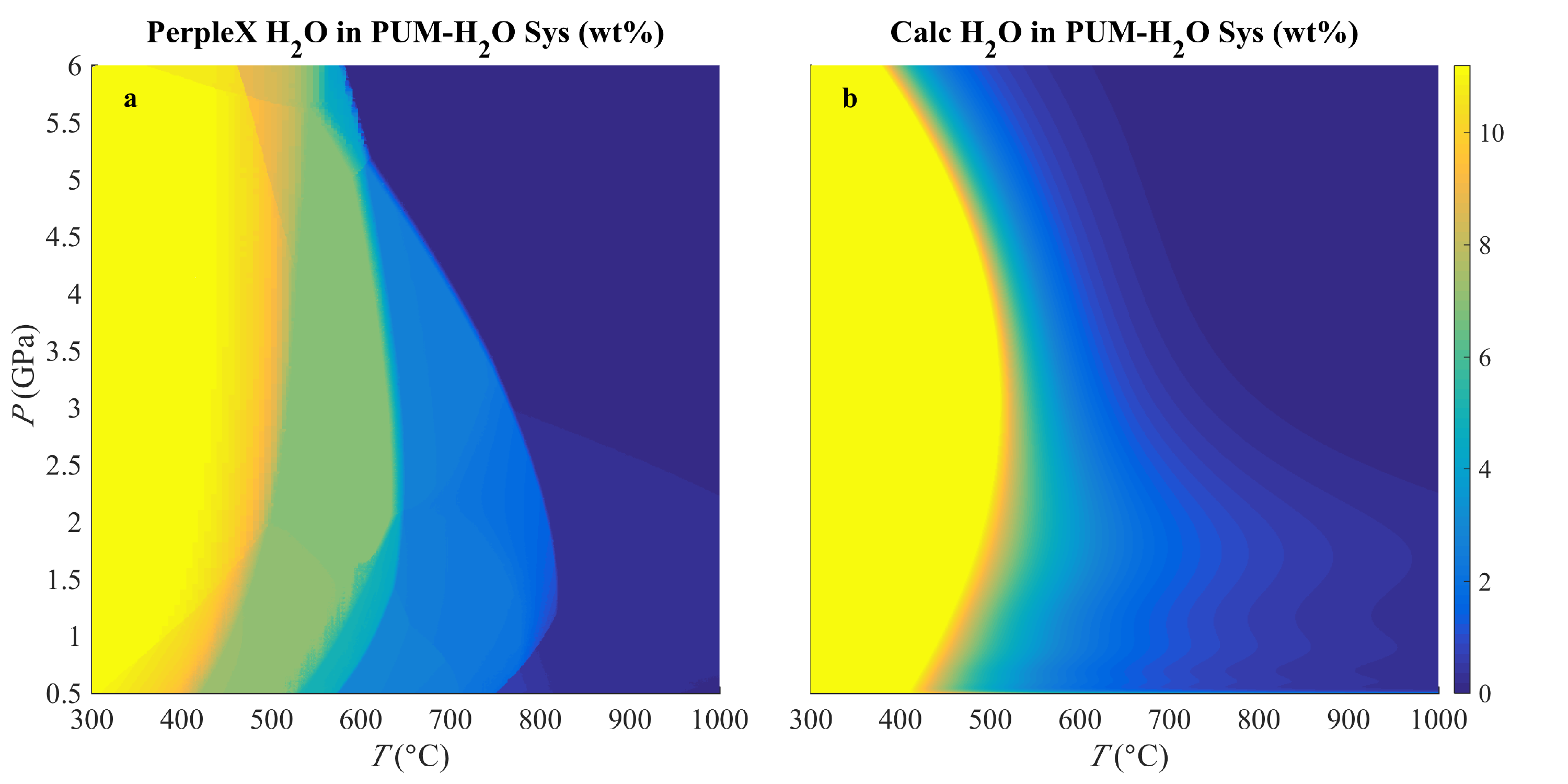} 
\caption{Comparison between the peridotite H$_2$O contents ($c_s^{\mathrm{H_2O}}$) derived from PerpleX (a) and our parameterization (b) for the peridotite--H$_2$O subsystem. ``PUM" in the titles is short for peridotite in the upper mantle, and the same as below. }
\label{fig: PUM-h2o}
\end{figure}

The volatile-free bulk composition for representative upper mantle is adopted from \citet{Hart:1986aa} and provided in Table \ref{tab: composition}. This composition is similar to the one for depleted mantle in \citet{Hacker:2008aa}, and is thus used for slab upper mantle composition residual to partial melting. The water content of this peridotite under H$_2$O-saturated condition is calculated by PerpleX and presented in Figure~\ref{fig: PUM-h2o}a. It is evident that there are three major dehydration reactions taking place with increasing temperature---breakdown of brucite, serpentine, and chlorite. Our parameterized result shown in Figure~\ref{fig: PUM-h2o}b captures the overall behavior of peridotite dehydration: the onset temperature of dehydration ($T_d^{ph}$) between $\sim$400--550 $^{\circ}$C approximates that of brucite decomposition; dewatering is more gradual at lower pressures ($<$ $\sim$3.5 GPa) because the three major dehydration reactions are more separated in temperature. Details of the parameterization are provided in Appendix \ref{apx:B}.   

\section{Simple Applications---Open System Effects} \label{sec:examples}
The parameterization performed above serves as a computational module that is interfaceable with reactive fluid flow modelling, under the assumption that local equilibrium holds during fluid--rock interaction (as in our companion paper, Part II). Although it ignores the details of specific dehydration and decarbonation reactions, the overall features of representative slab lithologies are captured. For example, released volatiles are H$_2$O rich close to the onset of devolatilization; more CO$_2$ comes into liquid phase with increasing temperature \citep{Molina:2000aa}. Since fluid flow renders the volatile-bearing system open, modelling of it entails thermodynamic computation that can accept evolving bulk composition and return the equilibrated thermodynamic state. Our parameterization meets this demand while being computationally efficient, and can thus simulate important open-system scenarios without needing repeated access to thermodynamic software (e.g. PerpleX) outside fluid flow models. To demonstrate this, we present its application to the two major open-system effects below.

\subsection{Effects of Fluid Removal (Fractionation)}
In open systems, the released H$_2$O and\slash or CO$_2$ will migrate through the system due to buoyancy and other forces, desiccating formerly volatile-bearing rocks. Differential H$_2$O and CO$_2$ loss will alter the H$_2$O\slash CO$_2$ ratio in rock residues, giving rise to chemical fractionation. Consider an extreme case where H$_2$O is preferentially lost from a gabbro such that the residual volatile species is only CO$_2$. The system evolves to the CO$_2$-only subsystem as in section \ref{sec:gab-co2}. As demonstrated earlier, the onset temperatures of decarbonation are higher than those of dehydration. Therefore chemical fractionation will lead to elevated onset temperatures of devolatilization and inhibit further volatile loss. 

\begin{figure}[h!]
\centering
\includegraphics[width=0.75\columnwidth, keepaspectratio]{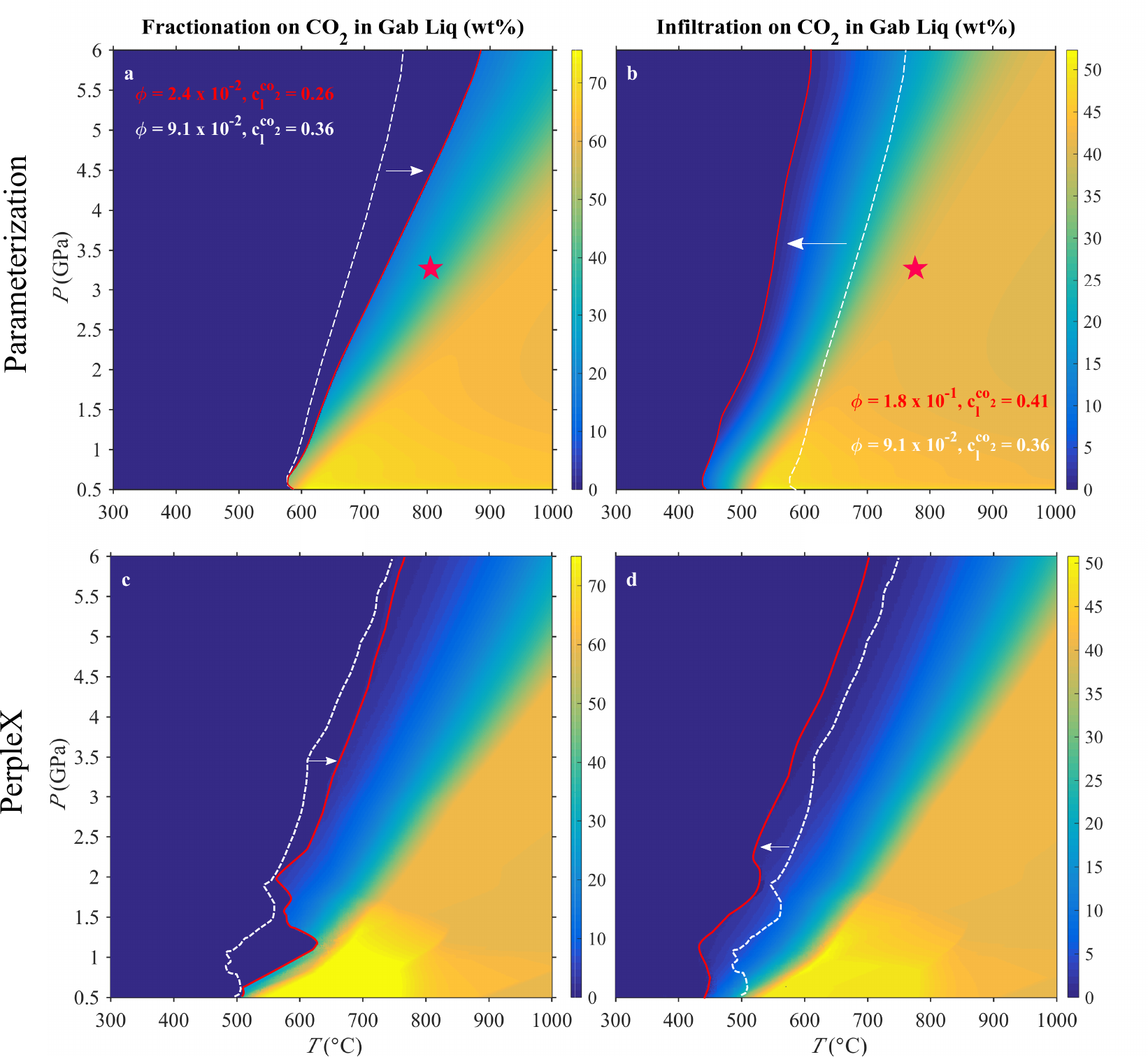} 
\caption{Illustration of the effects of fractionation (a and c) and infiltration (b and d). The upper panels show the results from our parameterization, whereas the lower panels show those from PerpleX, all based on the same bulk H$_2$O and CO$_2$ contents as detailed below. ``Gab liq" in the titles denote liquid phase equilibrated with gabbro. The white dashed lines depict the onset of devolatilization for the reference case of full gabbro--H$_2$O--CO$_2$ system with 2.58 wt\% bulk H$_2$O and 2.84 wt\% bulk CO$_2$ (section \ref{sec:gab-co2-h2o}). Red solid lines denote the onset of devolatilization after bulk H$_2$O \& CO$_2$ alterations by fractionation (panels a and c; with $c_{blk}^{\mathrm{H_2O}}$ = 1.57 wt\% and $c_{blk}^{\mathrm{CO_2}}$ = 2.30 wt\%) and by infiltration (panels b and d; with $c_{blk}^{\mathrm{H_2O}}$ = 4.13 wt\% and $c_{blk}^{\mathrm{CO_2}}$ = 2.79 wt\%). White arrows show the shift of onset temperatures from the reference case. Red stars mark the example pressure and temperature conditions used for comparison (see text). The corresponding porosity level and CO$_2$ weight fraction in liquid phase are given in the annotations.}
\label{fig: fracandinfi}
\end{figure}

An example is given in Figure \ref{fig: fracandinfi}a, where the white dashed line denotes the onset temperature of devolatlization for a representative gabbro containing 2.58 wt\% H$_2$O and 2.84 wt\% CO$_2$ (see Fig. \ref{fig: gab-h2o-co2}f). At 780 $^{\circ}$C and 3.25 GPa (star symbol in Fig. \ref{fig: fracandinfi}a), this bulk composition yields the equilibrated porosity $\mathrm{9.1 \times 10 ^{-2}}$ and liquid phase CO$_2$ mass fraction 0.36. The porosity level is equivalent to a liquid phase mass fraction $f=\mathrm{3.23 \times 10^{-2}}$, if densities of liquid and solid are taken to be 1000 kg m$^{-3}$ and 3000 kg m$^{-3}$, respectively. For illustrative purposes, assuming that half of the liquid phase leaves the bulk system due to fluid flow, then the renormalized bulk volatile content is:
\begin{equation} \label{eq:renorm}
c^{i'}_{blk} = \frac{c^i_{blk} - 0.5 f c^i_l}{1 - 0.5 f}, 
\end{equation}
where $i$ represents either H$_2$O or CO$_2$, and prime indicates renormalized values. The bulk H$_2$O content of 1.57 wt\% and CO$_2$ content of 2.30 wt\% after fractionation can be calculated from equation~\eqref{eq:renorm}. Figure \ref{fig: fracandinfi}a shows the new equilibration with the renormalized bulk volatile content, where the red line marks the new onset temperatures of devolatilization. It is evident that the chemical fractionation causes elevation of the onset temperatures of  devolatilization. At 780 $^{\circ}$C and 3.25 GPa, the newly equilibrated porosity is $\mathrm{2.4 \times 10 ^{-2}}$ and liquid phase CO$_2$ mass fraction is 0.26, both lower than the pre-fractionation levels, demonstrating the increased difficulty in devolatilization, especially decarbonation.

\subsection{Effects of Fluid Addition (Infiltration)}
Infiltration-driven decarbonation has been invoked as a potential mechanism of leaching CO$_2$ from subducting slabs to mantle wedge \citep[e.g.,][]{Kerrick:2001aa}. When H$_2$O-rich fluids infiltrate porous rocks, the pore fluid that was in equilibrium becomes diluted in terms of CO$_2$ concentration and thus out of equilibrium. To restore fluid--rock equilibrium, the rock will release CO$_2$ into the liquid phase to counterbalance the dilution. As infiltration proceeds, CO$_2$ is constantly lost from rocks, giving rise to infiltration-driven decarbonation. This idealized scenario obscures one subtlety associated with fluid infiltration---a process that is, to some extent, reverse to the fractionation discussed in the previous section. 

We highlight this subtlety through the following example. When a gabbroic rock is infiltrated by H$_2$O-rich fluids, the extra H$_2$O not only dilutes CO$_2$ in pore fluids, but also raises the bulk H$_2$O content in the fluid--rock system. Re-equilibration at elevated bulk H$_2$O content hence attempts not only to relieve the dilution via decarbonation reactions, but also to increase the fraction of coexisting liquid phase (i.e., porosity). Consider an extreme case where the infiltrating fluids are purely H$_2$O (e.g., from serpentinites), and the amount of added fluids is half that in the reference case of gabbro with 2.58 wt\% bulk H$_2$O and 2.84 wt\% CO$_2$. At 780 $^{\circ}$C and 3.25 GPa (red star in Fig. \ref{fig: fracandinfi}b), the coexisting liquid mass fraction before infiltration is $f=\mathrm{3.23 \times 10^{-2}}$, so the bulk volatile content after infiltration can be calculated by:
\begin{equation} \label{eq:renorm-infil}
c^{i'}_{blk} = \frac{c^i_{blk} + 0.5 f c^i_l}{1 + 0.5 f}, 
\end{equation}
where $i$ represents either H$_2$O or CO$_2$, and prime indicates renormalized values. With $c^{\mathrm{H_2O}}_l = 1$ and $c^{\mathrm{CO_2}}_l = 0$ for pure H$_2$O infiltration, the renormalized values are thus 4.13 wt\% bulk H$_2$O and 2.79 wt\% bulk CO$_2$ after infiltration. Figure \ref{fig: fracandinfi}b shows the CO$_2$ mass fraction in the liquid phase under the new equilibration. The white dashed line marks the onset of devolatilization before infiltration (same as in Fig. \ref{fig: fracandinfi}a), whereas the red solid line marks that after infiltration. It is clear that the elevated bulk H$_2$O content caused by infiltration shifts the onset of devolatilization to lower temperatures. Thus, when temperature and pressure are fixed, such a shift can enhance the extent of devolatilization. In particular, at 780 $^{\circ}$C and 3.25 GPa (red star in Fig. \ref{fig: fracandinfi}b), the infiltration not only raises porosity from $\mathrm{9.1 \times 10^{-2}}$ to $\mathrm{1.8 \times 10^{-1}}$, but also increases the liquid phase CO$_2$ content from 36 wt\% to 41 wt\%. Therefore, the potential of H$_2$O infiltration in driving decarbonation comes not only from dilution of coexisting liquid phase, but also from the increase of devolatilization extent.

Figure~\ref{fig: fracandinfi}c--d shows PerpleX calculations of the fractionation and infiltration effects in comparison with calculations using our parameterization (Fig.~\ref{fig: fracandinfi}a--b). In general, the shift of devolatilization onset temperatures is bigger in our parameterized results than in the PerpleX ones. For fractionation, the shift in the parameterization (Fig.~\ref{fig: fracandinfi}a) is up to 50 $^{\circ}$C higher than that in the PerpleX calculation (Fig.~\ref{fig: fracandinfi}c) above $\sim$3.5 GPa, but it decreases below $\sim$3.5 GPa. For infiltration, the $\sim$50 $^{\circ}$C reduction in the shift of devolatilization onset temperatures (Fig.~\ref{fig: fracandinfi}b \& d) is virtually constant throughout the entire pressure range considered. This disparity in the predicted fractionation and infiltration effects comes from the highly averaged treatment of the gabbro+H$_2$O subsystem. As shown in Figure~\ref{fig: gab-h2o}, there is a difference in the dehydration onset temperatures ($T_d^{gh}$) between the parameterized and PerpleX results, but such a difference is negligible for the gabbro+CO$_2$ subsystem (Fig.~\ref{fig: gab-co2}). When parameterizing the full system according to equations~\eqref{eqn:deltaT}--\eqref{eqn:Th2o}, this difference in the subsystem $T_d^{gh}$ is inherited, leading to the disparity between the top and bottom panels in Figure~\ref{fig: fracandinfi}. 

\section{Discussion} \label{sec:discussion}
The parameterization presented thus far can model open-system thermodynamic behaviors for a rock--H$_2$O--CO$_2$ system, where ``rock" represents typical lithologies in subducting slabs, i.e., sediments, basalts, gabbros, and peridotites. For each rock type, the parameterization predicts that CO$_2$ is released from the solid phase at higher temperature and lower pressure than H$_2$O, in accordance with experimental observations. Additionally, as dehydration and decarbonation in respective subsystems take place at different temperatures, the relative amounts of H$_2$O and CO$_2$ in bulk rocks control the onset temperature of overall devolatilization. The parameterization quantifies this compositional effect. These features enable efficient simulation of fractionation and infiltration during fluid flow, without the need to reproduce the full thermodynamics of devolatilization (i.e., compositional change of each mineral phase). As such, our approach balances model capability and complexity. Of course this balance means that some limitations are introduced, as compared with, e.g., PerpleX.

The first limitation is that the parameterization deals with systems that are open only to H$_2$O and CO$_2$. The effects of transport of other major, minor, and trace elements are not considered in the parameterization. For example, potassium is mobile in the presence of aqueous fluids \citep[see][for a review]{Ague:2014ab} and its loss can destablize phengite, a major H$_2$O-bearing mineral phase \citep{Schmidt:1996aa, Connolly18}, and therefore alter the H$_2$O partition coefficients (eq.~\eqref{eq:sub-partition}) in a fundamental way. Furthermore, as calcium and magnesium are the primary elements that stablize carbonates (e.g., calcite, magnesite), their mobilization can also considerably affect the CO$_2$ partition coefficients. In consequence, in the case of significant non-volatile element loss or gain \citep[e.g.][]{Philippot:1991aa, Ague:2014aa}, the accuracy of our parameterization deteriorates.

The second limitation of the current parameterization is the omission of partial melting. Given that sediments and basalts occupy the hottest region near slab surface and are more likely to melt, the limitation of our parameterization is associated with neglecting the partial melting of pelites and basalts, rather than gabbros or peridotites. Extensive experimental studies have been conducted to determine the solidus of carbonated pelites \citep[e.g.,][]{Thomsen:2008aa, Grassi:2011aa, Tsuno:2012aa, Tsuno:2012ab, Mann:2015aa} and carbonated basalts \citep[e.g.,][]{Yaxley:2004aa, Tsuno:2011aa, Tsuno:2012aa, Tsuno:2012ab, Dasgupta:2004aa}. Regarding carbonated pelites, \citet{Thomsen:2008aa} determined the solidus to be 900~$^{\circ}$C to 1070~$^{\circ}$C at pressures from 2.4 to 5.0~GPa. In a subsequent experiment on a sample with similar H$_2$O and CO$_2$ content, \citet{Tsuno:2012aa} determined a melting temperature between 800~$^{\circ}$C and 850~$^{\circ}$C at 3.0~GPa. As reviewed by \citet{Mann:2015aa}, the wet solidus for pelites ranges from 745~$^{\circ}$C to 860~$^{\circ}$C at pressures from 3.0 to 4.5~GPa in the presence of CO$_2$, but the fluid-absent solidus is from 890~$^{\circ}$C to 1040~$^{\circ}$C. These studies suggest that carbonated silicate melting or carbonatite melting are unrealistic for subducting sediments due to the low slab geotherms, unless H$_2$O saturation is realized near the wet solidus of carbonated sediments. Hence, our current parameterization for sediment devolatilization is applicable to the scenarios where sediments are either H$_2$O undersaturated or H$_2$O saturated but at temperatures below the wet solidus. On the other hand, the solidus of carbonated basalts (eclogites) was investigated by \citet{Dasgupta:2004aa}, \citet{Yaxley:2004aa}, and \citet{Tsuno:2011aa}, and was shown to exceed the maximum temperatures that can be reached within subducting slabs. This supports our omission of basalt partial melting in the parameterization. Thus, neglect of partial melting in our parameterization is valid in most cases. In rare cases where slab melting occurs as inferred from surface adakitic magmatism \citep{Drummond:1996aa}, the current parameterization becomes inappropriate.

The third limitation is the overestimation of open-system effects caused by fluid removal and addition, as discussed in section \ref{sec:examples}. This limitation is inherent in our parameterization, and not relevant to PerpleX. Qualitatively, fractionation tends to inhibit devolatilization whereas infiltration tends to promote it. In terms of modelling coupled H$_2$O \& CO$_2$ transport in subducting slabs, these effects are likely to counterbalance each other. Quantitatively, however, it is uncertain to what extent this counterbalance is effective. A future improvement might look for alternative functional forms other than the quadratic one (eq.~\eqref{eqn:Tco2}--\eqref{eqn:Th2o}) to achieve better predictions of onset temperatures of devolatilization. Nevertheless, compared with the conventional approaches that are fully based on PerpleX but ignore open-system behaviors, the current parameterization allows efficient treatment of open-system behaviors. With the recent increasing recognition of carbon transport via dissolved ionic species in subduction zones \citep[e.g.,][]{Frezzotti:2011aa}, open-system effects are thought to play a dominant role in carbon transport. Therefore, a thermodynamic model capable of dealing with fractionation and infiltration seems highly desirable. The parameterization approach would be particularly advantageous when ionic species are included; in particular, the first limitation above is overcome.

\section{Conclusion}
In light of considerable uncertainties regarding the spatio-temporal pattern of fluid flow and the equilibrium state in the downwelling slabs, a balance between model capability and complexity should be maintained when assessing the slab H$_2$O and CO$_2$ budget during subduction. We present a thermodynamic parameterization for the devolatilization of representative lithologies in slabs comprising sediments, basalts, gabbros, and peridotites. This parameterization achieves an appropriate balance in that it captures the leading-order features of coupled decarbonation and dehydration, while it smooths over the details of specific mineral reactions, either continuous or discontinuous. The first captured feature is that equilibrated fluids are increasingly CO$_2$-rich with elevating temperature and reducing pressure, and the second is that increasing the CO$_2$\slash{}H$_2$O ratio in bulk rocks increases onset temperatures of devolatilization and thus inhibits overall devolatilization. With these two features, the parameterization is able to simulate the two open-system behaviors that significantly affect slab H$_2$O and CO$_2$ transport: fractionation and infiltration. Nonetheless, the current parameterization is limited to the scenarios in which metasomatism does not considerably alter the bulk rock composition with regard to non-volatile elements and the subducting slab does not partially melt. Within these limits, the parameterization can be efficiently coupled to reactive fluid flow modelling for the study of slab H$_2$O and CO$_2$ budget during subduction, the results of which are presented in the companion paper Part II. 

\appendix
\section{Derivation of the Formalism for Parameterization} \label{apx:A}
For volatile $i$ (H$_2$O or CO$_2$) that reaches equilibrated partition between solid and liquid phases:
\begin{equation}
\mu^i_{s} = \mu^i_{l}, 
\end{equation}
namely, the chemical potential ($\mu$) of this volatile component are equal between the two phases. Expanding on the thermodynamic state of pure material of component $i$:
\begin{equation} \label{eq:expansion}
\mu^i_{s0} (P, T) + R T \ln a^i_s = \mu^i_{l0} (P, T) + R T \ln a^i_l, 
\end{equation}
where $a^i_s$ and $a^i_l$ are the activity of component $i$ in solid and liquid phases, respectively, and the subscript ``$0$" denotes the chemical potential of pure material. Equation~\eqref{eq:expansion} can be rearranged as:
\begin{equation} \label{eq:govern1}
\ln \frac{a^i_s}{a^i_l} = \frac{1}{R T} \left[ \mu^i_{l0} (P, T) -  \mu^i_{s0} (P, T) \right], 
\end{equation}
where the right hand side is a function of only pressure ($P$) and temperature ($T$), not composition. In the ideal solution theory, dissolution of other components has no effects on the component $i$, indicating that the activity ratio of $a^i_s$ over $a^i_l$ is approximately that of respective concentrations ($c$), namely:
\begin{equation} \label{eq:ideal-K}
\ln \frac{a^i_s}{a^i_l} \approx \ln \frac{c^i_s}{c^i_l} = \ln K^i, 
\end{equation}
where $K^i$ is the partition coefficient of component $i$. In the case of non-ideal solution, we have:
\begin{equation}
\ln \frac{a^i_s}{a^i_l} = \ln \frac{\gamma^i_s c^i_s}{\gamma^i_l c^i_l}, 
\end{equation}
where $\gamma^i_s$ and $\gamma^i_l$ are the activity coefficients for component $i$ in solid and liquid phases, respectively. If it is further assumed that volatile concentrations in solid phase ($c^i_s$) is small such that Henry's Law of dilute solution applies, 
\begin{equation} \label{eq:nonideal-K}
\ln \frac{a^i_s}{a^i_l} = \ln \frac{\gamma_0 c^i_s}{\gamma^i_l c^i_l} = \ln \frac{K^i}{\gamma^i}, 
\end{equation}
where $\gamma^i = \gamma^i_l / \gamma_0$, and $\gamma_0$ is a constant according to Henry's Law.

The right hand side of equation~\eqref{eq:govern1} can be transformed to:
\begin{equation} \label{eq:expansion-rhs}
\frac{1}{R T} \left[ \mu^i_{l0} (P, T) -  \mu^i_{s0} (P, T) \right] = \frac{1}{R T} \left[ \Delta H(P, T)  - T \Delta S(P, T)  \right], 
\end{equation}
where $\Delta H$ and $\Delta S$ are the enthalpy and entropy changes associated with an effective reaction---the reaction that involves the conversion from solid to liquid state for material consisted of purely component $i$. Within the pressure and temperature range a subducting slab undergoes, there are multiple dehydration or decarbonation reactions. If we approximate $\Delta H$ and $\Delta S$ by their average values corresponding to the multiple reactions ($\Delta H$ and $\Delta S$ represent respectively the enthalpy and entropy changes of an effectively averaged dehydration ($i = \mathrm{H_2O}$) or decarbonation ($i = \mathrm{CO_2}$) reaction), then

\begin{equation} \label{eq:appro}
\Delta H (P, T) \approx \Delta H^i (P) = T^i_d (P) \Delta S^i(P) \approx T^i_d (P) \Delta S (P, T), 
\end{equation}
where $T^i_d (P)$ is the onset temperature of the averaged dehydration or decarbonation. Note that since the onset of average reaction is univariant and corresponds to a line in $P$-$T$ space, that is, $\Delta H^i(T_d^i, P) = \Delta H^i(T_d^i(P), P) = \Delta H^i(P)$ and similarly for $\Delta S^i(T_d^i, P)$, the $\Delta H^i$ and $\Delta S^i$ in equation~\eqref{eq:appro} are written as pressure dependent only. Substituting equation~\eqref{eq:appro} into equation~\eqref{eq:expansion-rhs}, we get:
\begin{equation} \label{eq:approx-rhs}
\frac{1}{R T} \left[ \mu^i_{l0} (P, T) -  \mu^i_{s0} (P, T) \right] \approx \frac{\Delta H^i (P)}{R} \left[ \frac{1}{T} - \frac{1}{T^i_d} \right]. 
\end{equation}

To conclude, for ideal solutions, equations~\eqref{eq:govern1}, \eqref{eq:ideal-K} and \eqref{eq:approx-rhs} lead to:
\begin{equation} 
K^i \approx \exp \left[ \frac{\Delta H^i (P)}{R} \left( \frac{1}{T} - \frac{1}{T^i_d} \right) \right], 
\end{equation}
whereas for non-ideal solutions, equations~\eqref{eq:govern1}, \eqref{eq:nonideal-K} and \eqref{eq:approx-rhs} lead to:
\begin{equation} 
\frac{K^i}{\gamma^i} \approx \exp \left[ \frac{\Delta H^i (P)}{R} \left( \frac{1}{T} - \frac{1}{T^i_d} \right) \right]. 
\end{equation}
Let $L_R (P) =  \Delta H^i (P)/R $ and adopt $\mathcal{K}^i$ for the partition coefficients in the non-ideal cases:
\begin{equation} \label{eq:govern-ideal}
K^i \approx \exp \left[ L_R \left( \frac{1}{T} - \frac{1}{T^i_d} \right) \right], 
\end{equation}
\begin{equation} \label{eq:govern-nonideal}
\mathcal{K}^i \approx \gamma^i \exp \left[ L_R \left( \frac{1}{T} - \frac{1}{T^i_d} \right) \right]. 
\end{equation}
If a common factor of saturation content $c^i_{sat}$ is further singled out from equations~\eqref{eq:govern-ideal} and \eqref{eq:govern-nonideal}, then the formalism of equations~\eqref{eq:sub-partition} and \eqref{eqn:non-ideal-partition} respectively in the main text is recovered.

\section{Polynomials Fitted in the Parameterization} \label{apx:B}
\subsection{MORB}
\subsubsection{MORB--H$_2$O Subsystem}
For the MORB--H$_2$O subsystem, we found the following polynomials best fit the PerpleX-derived data:
\begin{equation}
\ln(c_{sat}^{mh}(P)) = a_0 P^3 + a_1 P^2 + a_2 P + a_3, 
\end{equation}
\begin{equation}
\ln(L_R^{mh}(P)) = b_0 /P^4 + b_1 /P^3 + b_2 /P^2 + b_3 /P + b_4,
\end{equation}
\begin{equation}
T_d^{mh}(P) = c_0 P^2 + c_1 P + c_2,
\end{equation}
and the polynomial coefficients are provided in Table \ref{tab:morb}.

 \begin{table}[h!]
 \caption{Regression Results for MORB--H$_2$O--CO$_2$ System}
 \label{tab:morb}
 \centering
 \begin{ruledtabular}
 \begin{tabular}{l c c c c c c c c}
 
     & \multicolumn{6}{c@{\hspace{0.5cm}}|}{H$_2$O} & \multicolumn{2}{c}{CO$_2$} \\
 \hline
   $c_{sat}$ & -- & -- & 0.0102725 & $-$0.115390 & 0.324452 & \multicolumn{1}{c@{\hspace{0.5cm}}|}{1.41588} & -- & 19.0456 \\
   $L_R$ & -- & $-$1.78177 & 7.50871 & $-$10.4840 & 5.19725 & \multicolumn{1}{c@{\hspace{0.5cm}}|}{7.96365} & 0.505130 & 10.6010 \\
   $T_d$ & -- & -- & -- & $-$3.81280 & 22.7809 & \multicolumn{1}{c@{\hspace{0.5cm}}|}{638.049} & 116.081 & 826.222 \\    
\hline
   $W_{\mathrm{H_2O}}$ & -- & -- & -- & $-$0.05546 & 0.8003 & \multicolumn{1}{c@{\hspace{0.5cm}}}{$-$3.595} & 5.155 & 10.02 \\ 
   $W_{\mathrm{CO_2}}$ & $-$0.001217 & 0.03207 & $-$0.3466 & 1.98 & $-$6.419 & \multicolumn{1}{c@{\hspace{0.5cm}}}{11.75} & $-$11.08 & 15.06 \\       
 
 \end{tabular}
 \end{ruledtabular}
 \end{table}

\subsubsection{MORB--CO$_2$ Subsystem}
Similar to the gabbro--CO$_2$ subsystem, we found the following polynomials best fit the PerpleX-derived data of the MORB--CO$_2$ subsystem:
\begin{equation}
c_{sat}^{mc}(P) = a_0, 
\end{equation}
\begin{equation}
\ln(L_R^{mc}(P)) = b_0 /P + b_1, 
\end{equation}
\begin{equation}
T_d^{mc}(P) = c_0 P + c_1,
\end{equation}
where values of coefficients are listed in Table \ref{tab:morb}.  

\subsubsection{MORB--H$_2$O--CO$_2$ Full System}
\begin{figure}[h!]
\centering
\includegraphics[width=0.8\columnwidth, keepaspectratio]{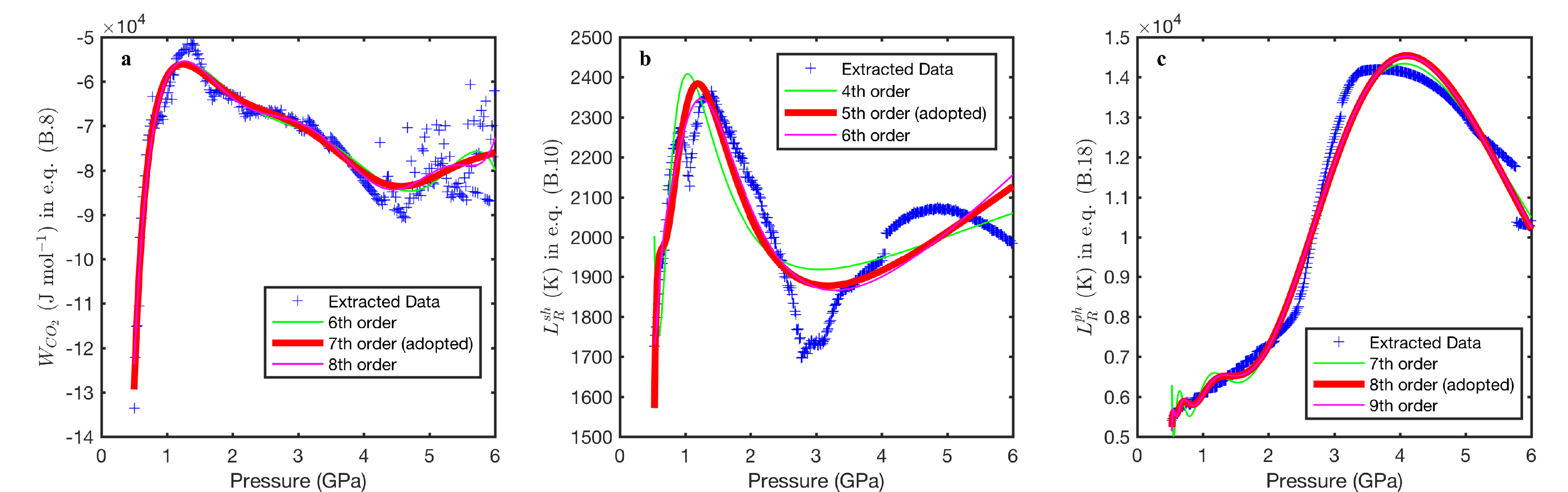} 
\caption{Illustration for the quality of high-order polynomial fitting for $W_{CO_2}$, $L_R^{sh}$, and $L_R^{ph}$ from equations~\eqref{eq:b8}, \eqref{eq:b10}, and \eqref{eq:ph-lr}, respectively. The thickened solid red lines denotes the polynomials adopted. }
\label{fig: regression_appendix}
\end{figure}

In fitting the data of $W_{\mathrm{H_2O}}$ and $W_{\mathrm{CO_2}}$ derived from calibrating against PerpleX calculation, we experiment with polynomials of pressure from low to high orders, and the following forms turn out to best fit the data:
\begin{equation}
\ln(-W_{\mathrm{H_2O}}) = d_0 P^4 + d_1 P^3 + d_2 P^2 + d_3 P + d_4,
\end{equation}
\begin{equation} \label{eq:b8}
\ln(-W_{\mathrm{CO_2}}) = e_0 P^7 + e_1 P^6 + e_2 P^5 + e_3 P^4 + e_4 P^3 + e_5 P^2 + e_6 P + e_7.
\end{equation}
The fitted coefficients are provided in Table \ref{tab:morb}. For fitting of polynomials higher than the fifth order (e.g., equation~\eqref{eq:b8}), we illustrate the quality of regression in Figure~\ref{fig: regression_appendix}, which additionally includes those for equations~\eqref{eq:b10} and \eqref{eq:ph-lr}.  

\subsection{Sediment}
\subsubsection{Sediment--H$_2$O Subsystem}
For the H$_2$O-only subsystem, using the volatile-free bulk composition for sediments in Table \ref{tab: composition}, values of $c_{sat}^{\mathrm{H_2O}}$, $L_R$, and $T_d^{\mathrm{H_2O}}$ can be extracted from a PerpleX run assuming H$_2$O saturation. The following polynomial forms of pressure turn out to best fit the PerpleX-derived data, and the fitted coefficients are provided in Table \ref{tab:sed}.
\begin{equation}
\ln(c_{sat}^{sh}(P)) = a_0 (\log P)^2 + a_1 (\log P) + a_2, 
\end{equation}
\begin{equation} \label{eq:b10}
\ln(L_R^{sh}(P)) = b_0 /P^5 + b_1 /P^4 + b_2 /P^3 + b_3 /P^2 + b_4 /P + b_5,
\end{equation}
\begin{equation}
T_d^{sh}(P) = c_0 P^3 + c_1 P^2 + c_2 P + c_3.
\end{equation}

\begin{table}[h!]
 \caption{Regression Results for Sediment--H$_2$O--CO$_2$ System}
 \label{tab:sed}
 \centering
 \begin{ruledtabular}
 \begin{tabular}{l c c c c c c c c}
 
     & \multicolumn{6}{c@{\hspace{0.5cm}}|}{H$_2$O} & \multicolumn{2}{c}{CO$_2$} \\
 \hline
   $c_{sat}$ & -- & -- & -- & $-$0.150662 & 0.301807 & \multicolumn{1}{c@{\hspace{0.5cm}}|}{1.01867} & -- & 10.5923 \\
   $L_R$ & $-$2.03283 & 10.8186 & $-$21.2119 & 18.3351 & $-$6.48711 & \multicolumn{1}{c@{\hspace{0.5cm}}|}{8.32459} & 0.0974525 & 10.9734 \\
   $T_d$ & -- & -- & 2.83277 & $-$24.7593 & 85.9090 & \multicolumn{1}{c@{\hspace{0.5cm}}|}{524.898} & 92.5788 & 874.392 \\    
\hline
   $W_{\mathrm{H_2O}}$ & -- & -- & -- & -- & -- & \multicolumn{1}{c@{\hspace{0.5cm}}}{--} & 0.1045 & 12.53 \\ 
   $W_{\mathrm{CO_2}}$ & -- & -- & -- & -- & -- & \multicolumn{1}{c@{\hspace{0.5cm}}}{--} & -- & $-\mathrm{10^4}$ \\       
 
 \end{tabular}
 \end{ruledtabular}
\end{table}

\subsubsection{Sediment--CO$_2$ Subsystem}
Due to the limited varieties of carbonate mineral phases, the polynomial forms used for fitting the sediment--CO$_2$ subsystem are the same as in earlier subsystems:
\begin{equation}
c_{sat}^{sc}(P) = a_0, 
\end{equation}
\begin{equation}
\ln(L_R^{sc}(P)) = b_0 /P + b_1, 
\end{equation}
\begin{equation}
T_d^{sc}(P) = c_0 P + c_1,
\end{equation}
where the fitted coefficients are listed in Table \ref{tab:sed}. 

\subsubsection{Sediment--H$_2$O--CO$_2$ Full System}
In the full sediment--H$_2$O--CO$_2$ system, we found the following polynomials fitted for $W_{\mathrm{H_2O}}$ and $W_{\mathrm{CO_2}}$ best reproduce the result from PerpleX calculation, and the coefficients are listed in Table \ref{tab:sed}.
\begin{equation}
\ln(-W_{\mathrm{H_2O}}) = d_0 P + d_1,
\end{equation}
\begin{equation}
\ln(-W_{\mathrm{CO_2}}) = e_0.
\end{equation}

\begin{table}[h!]
\caption{Regression Results for Peridotite--H$_2$O Subsystem}
\label{tab:pum}
\centering
\begin{ruledtabular}
\begin{tabular}{l c c c c c c c c c}

     & \multicolumn{9}{c}{H$_2$O}  \\
\hline
   $c_{sat}$ & -- & -- & -- & -- & -- & -- & -- & 0.00115628 & 2.42179 \\
   $L_R$ & $-$19.0609 & 168.983 & $-$630.032 & 1281.84 & $-$1543.14 & 1111.88 & $-$459.142 & 95.4143 & 1.97246 \\
   $T_d$ & -- & -- & -- & -- & -- & -- & $-$15.4627 & 94.9716 & 636.603 \\    

\end{tabular}
\end{ruledtabular}
\end{table}

\subsection{Peridotite of Upper Mantle}
The same procedure as for other subsystems is adopted to derive $c_{sat}^{\mathrm{H_2O}}$, $T_d^{\mathrm{H_2O}}$, and $L_R$ from PerpleX using the volatile-free bulk composition for peridotite (Table \ref{tab: composition}). In fitting the data as polynomial functions of pressure, we found the following achieve the best fit:
\begin{equation}
\ln(c_{sat}^{ph}(P)) = a_0 P + a_1, 
\end{equation}
\begin{equation} \label{eq:ph-lr}
\ln(L_R^{ph}(P)) = b_0 /P^8 + b_1 /P^7 + b_2 /P^6 + b_3 /P^5 + b_4 /P^4 + b_5/P^3 + b_6/P^2 + b_7/P + b_8,
\end{equation}
\begin{equation}
T_d^{ph}(P) = c_0 P^2 + c_1 P + c_2 .
\end{equation}
Regressed coefficients are listed in Table \ref{tab:pum} and comparison between PerpleX and the parameterization is given in Figure \ref{fig: PUM-h2o}. Note that equation~\eqref{eq:ph-lr} is 8th order in $P^{-1}$, this does not collapse our parameterization at high peressures because $P^{-1}$ turns negligibly small and it is the lower-order terms that matter. Nonetheless, these high-order terms are retained to fine-tune the variation of dehydration rate under low pressure conditions, as reflected by the variation of H$_2$O content with temperature below $\sim$2.5 GPa in Figure \ref{fig: PUM-h2o}a.

\clearpage

\begin{acknowledgments}
We thank Ikuko Wada and an anonymous reviewer for their comments that improve this manuscript. The authors thank the Isaac Newton Institute for Mathematical Sciences for holding the Melt in the Mantle program sponsored by EPSRC Grant Number EP/K032208/1. Support from Deep Carbon Observatory funded by the Sloan Foundation is acknowledged. M.T. acknowledges the Royal Society Newton International Fellowship (NF150745). D.R.J acknowledges research funding through the NERC Consortium grant NE/M000427/1 and NERC Standard grant NE/I026995/1. This project has also received funding from the European Research Council (ERC) under the European Union's Horizon 2020 research and innovation programme (grant agreement n$^{\circ}$ 772255). This contribution is about numerical modelling, so it does not depend on experimental or field data. Relevant data and equations for reproducing the model results are already contained in the text. Nonetheless, we provide an example code on the usage of the thermodynamic parameterization: \url{https://bitbucket.org/meng_tian/example_code_thermo_module/src/master/}.
\end{acknowledgments}
\bibliography{manuscriptref}

\end{document}